\documentclass[]{article}

\usepackage{subeqn}
\def\thesubequation{\themainequation\roman{equation}}

\usepackage[dvips]{graphicx}

\usepackage{paralist}
\usepackage{calrsfs}
\usepackage{oldgerm}
\usepackage{amssymb,amsbsy}
\usepackage{epic,eepic}
\usepackage{texdraw}
\usepackage{dsfont}
\usepackage{latexsym}
\usepackage{ntheorem}

\usepackage{accents}
\DeclareMathAccent{\wtilde}{\mathord}{largesymbols}{"65}
\DeclareMathAccent{\what}{\mathord}{largesymbols}{"62}

\makeatletter
\def\wb{\accentset{{\cc@style\underline{\mskip10mu}}}}
\makeatother

\newcommand{\be}{\begin{equation}}
\newcommand{\ee}{\end{equation}}
\newcommand{\bdm}{\begin{displaymath}}
\newcommand{\edm}{\end{displaymath}}

\theoremstyle{break}

\newtheorem{lm}{Lemma}[section]

\newtheorem{thm}{Proposition}[section]

\theorembodyfont{\upshape}

\newtheorem{rem}{Remark}[section]
\newtheorem*{pr}{Proof}

\makeatother

\begin{document}

\title{Continuous symmetric reductions of the Adler--Bobenko--Suris equations}

\author{Tsoubelis D and Xenitidis P \\
Department of Mathematics, University of Patras, 265 00 Patras, Greece\\}

\maketitle

E-mail : xeniti@math.upatras.gr

\begin{abstract}
Continuously symmetric solutions of the Adler-Bobenko-Suris class of discrete integrable equations are presented. Initially defined by their invariance under the action of both of the extended three point generalized symmetries admitted by the corresponding equations, these solutions are shown to be determined by an integrable system of partial differential equations. The connection of this system to the Nijhoff-Hone-Joshi ``generating partial differential equations'' is established and an auto-B{\"a}cklund transformation and a Lax pair for it are constructed. Applied to the {\it H1} and ${\it Q1}_{\delta=0}$  members of the Adler-Bobenko-Suris family, the method of continuously symmetric reductions yields explicit solutions determined by the Painlev{\'e} trancendents. 
\end{abstract}

\section{Introduction}

The study of integrable discrete systems has a long history going back to work in the late seventies and early eighties \cite{AblLad,Hirota,NQC,QNCL}. At this point, it is acknowledged that, most of the well known integrable discrete systems are characterized by their ``multidimensional consistency''. This means that, such a system may be imposed in a consistent way in a multidimensional space. This property seems to incorporate automatically two integrability aspects of this kind of systems, in the following sense: Multidimensional consistency allows one to derive algorithmically a B{\"a}cklund transformation, as well as, a Lax pair for the difference equations under consideration \cite{BobSuris,Nij1,Xen}.

Recently, Adler, Bobenko and Suris (ABS) classified the scalar lattice equations which are multidimensionally consistent and possess the symmetries of the square and the tetrahedron property, as well \cite{ABS}. Subsequently, they classified the lattice equations having the consistency property in a more general framework \cite{ABS1}. 

The equations covered by the ABS classification \cite{ABS} have already attracted the interest of many investigators and several results pertaining to them have already been published, including exact solutions \cite{AHN1,AHN2}, B{\"a}cklund transformations \cite{Atk}, symmetries \cite{TTX,RHsym,levi-petr1,levi-petr2} and conservation laws \cite{RHcons}.

In this paper, we focus on the symmetry properties of the ABS equations and show how a particular class of reductions provide a natural interplay between them and certain non-autonomous systems of partial differential equations. The means to explore this link is provided by the pair of extended three-point generalized symmetries admitted by the equations of the ABS class \cite{TTX}. 

More specifically, we study the {\it{continuously invariant solutions}} of the systems under consideration. We use the term ``continuously invariant solutions'' for the solutions that remain invariant under the action of both of the extended three-point generalized symmetries admitted by the corresponding equation. We show that these solutions are determined by a system of differential--difference equations, which involves six values of the unknown function, $u$. The elimination of three of these values leads to an equivalent system of partial differential equations, $\Sigma[u]$, which involves the remaining values of the dependent variable. 

Among the other advantages offered by the general framework of the continuously invariant solutions developed in this paper is the fact that it allows us to derive easily some of the integrability properties of $\Sigma[u]$. In particular, it enables us to construct an auto-B{\"a}cklund transformation for this system, as well as, a Lax pair.

The implementation of this general framework to the equations {\it H1} and ${\it Q1}_{\delta=0}$ of the ABS family leads to explicit solutions, constructed using symmetry reductions of the corresponding $\Sigma[u]$ systems. These solutions are determined by quadratures from the continuous Painlev{\'e} V and VI equations, but may also be regarded as resulting from reductions which lead to discrete Painlev{\'e} equations \cite{TTX,NijP6}. In this fashion, a new connection between discrete and continuous versions of the Painlev{\'e} equations is revealed.

Another important aspect of system $\Sigma[u]$ is that it leads to  what has been termed as {\it{generating partial differential equations}}. The archetypical example of such equations is the {\it{regular partial differential equation}} (RPDE), introduced by Nijhoff, Hone and Joshi in \cite{NHJ}. These authors showed that the RPDE, which encodes the entire hierarchy of the Korteweg - de Vries (KdV) equation, is related to equation {\it H1} of the ABS family. In the present paper the above result is rederived, but by a completely different method, which also allows its immediate generalization. Specifically, we show that, not only {\it H1}, but also {\it H2}, {\it H3} and {\it Q1} are related to the RPDE, and establish this relation in a systematic fashion, using the properties of the corresponding $\Sigma[u]$.

The present paper is organized as follows. In Section 2, we first introduce the notation used in the sections that follow. Then, we present the main characteristics of a wider class of lattice equations containing all the members of the ABS family, along with an auto-B{\"a}cklund transformation, ${\mathds{B}}_d$, for each member of the latter.

Section 3 deals with the solutions of the equations of ABS family which remain invariant under the action of the two extended three point generalized symmetries admitted by the above equations. These solutions are determined by a system of differential--difference equations which we prove to be equivalent to the integrable system $\Sigma[u]$. In the same section we prove that, the class of continuously invariant solutions is closed under the B{\"a}cklund transformation ${\mathds{B}}_d$. Exploiting this result, we derive two items revealing the integrability of system $\Sigma[u]$ itself, namely an auto-B{\"a}cklund transformation and a Lax pair.

Sections 4 and 5 contain the application of the general results of Section 3 to the ABS equations {\it H1} and ${\it Q1}_{\delta=0}$. Specifically, we construct symmetry reductions of the corresponding $\Sigma[u]$ systems, in terms of which, explicit solutions of the above equations are determined.

Section 6 deals with generating partial differential equations and the detailed analysis of system $\Sigma[u]$ corresponding to equations {\it H1}--{\it H3} and {\it Q1} is presented. In particular, we show that systems $\Sigma[u]$ for {\it H1}, {\it H2} and {\it Q1} are related, through a contact transformation, to RPDE. Also, we derive the connection of $\Sigma[u]$ for {\it H3} to RPDE. Finally, we relate our results to the ones of \cite{NHJ}, where the connection of {\it H1}, {\it H3}${}_{\delta=0}$ and {\it Q1}${}_{\delta=0}$ to RPDE was presented from a different point of view.

The concluding section contains an overall evaluation of the presented results and various perspectives.

\section{Notation and the Adler-Bobenko-Suris equations}

We first introduce the notation that will be used in what follows. In addition, we present those properties of the ABS equations that will be used in the next sections. 

\begin{figure}[h]
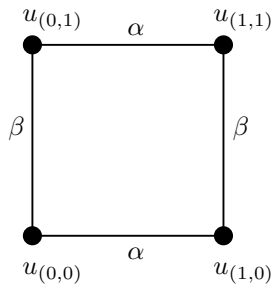

\centertexdraw{ \setunitscale 0.5
\linewd 0.02 \arrowheadtype t:F 
\htext(0 0.5) {\phantom{T}}
\move (-1 -2) \lvec (1 -2) 
\move(-1 -2) \lvec (-1 0) \move(1 -2) \lvec (1 0) \move(-1 0) \lvec(1 0)
\move (1 -2) \fcir f:0.0 r:0.1 \move (-1 -2) \fcir f:0.0 r:0.1
 \move (-1 0) \fcir f:0.0 r:0.1 \move (1 0) \fcir f:0.0 r:0.1  
\htext (-1.1 -2.5) {$u_{(0,0)}$} \htext (.9 -2.5) {$u_{(1,0)}$} \htext (0 -2.25) {$\alpha$}
\htext (-1.1 .15) {$u_{(0,1)}$} \htext (.9 .15) {$u_{(1,1)}$} \htext (0 .1) {$\alpha$}
\htext (-1.25 -1) {$\beta$} \htext (1.1 -1) {$\beta$}}
\caption{{\em{An elementary quadrilateral}}} \label{fig:quad}
\end{figure}

A partial difference equation is a functional relation among the values of a function $u : {\mathds{Z}} \times {\mathds{Z}} \rightarrow {\mathds{C}}$ at various points of the lattice, which may also involve the independent variables $n$, $m$ and the lattice spacings $\alpha$, $\beta$, see Figure \ref{fig:quad}, i.e. a relation of the form
\be {\cal E}(n,m,u_{(0,0)},u_{(1,0)},u_{(0,1)},\ldots;\alpha,\beta)\,=\,0\,. \label{gendisceq} \ee
In this relation, $u_{(i,j)}$ denotes the value of the function $u$ at the lattice point $(n+i,m+j)$, e.g.
$$u_{(0,0)}\,=\,u(n,m)\,,\quad u_{(1,0)}\,=\,u(n+1,m)\,,\quad u_{(0,1)}\,=\,u(n,m+1)\,,$$
and this is the notation that will be used for the values of the function $u$ from now on. 

The analysis of such equations is facilitated by the introduction of two translation operators acting on functions on ${\mathds{Z}}^2$, defined by 
$$\left( \mathcal{S}_n^{(k)} u \right)_{(0,0)} = u_{(k,0)}\,,\quad \left( \mathcal{S}_m^{(k)} u\right)_{(0,0)} = u_{(0,k)}\,,\quad {\mbox{where}}\,\, k \in \mathds{Z} \,,$$
respectively.

The equations of the ABS family belong to a wider class which contains all the equations of the form
\begin{equation}
Q(u_{(0,0)},u_{(1,0)},u_{(0,1)},u_{(1,1)};\alpha,\beta) \,=\, 0\,, \label{eq:genform}
\end{equation}
where the function $Q$ satisfies the following requirements:
\begin{enumerate}[1.]
\item It does not depend explicitly on the discrete variables $n$, $m$.
\item It is affine linear and depends explicitly on the four values of the unknown function $u$, i.e. 
$$\partial_{u_{(i,j)}} Q(u_{(0,0)},u_{(1,0)},u_{(0,1)},u_{(1,1)};\alpha,\beta) \, \ne \, 0$$
and 
$$\partial_{u_{(i,j)}}^2 Q(u_{(0,0)},u_{(1,0)},u_{(0,1)},u_{(1,1)};\alpha,\beta)\, =\, 0\,,$$
where $i$, $j$ = 0, 1.
\item It possesses the symmetries of the square (${\mathrm{D}}_4$-symmetry):
\begin{eqnarray*}
 Q(u_{(0,0)},u_{(1,0)},u_{(0,1)},u_{(1,1)};\alpha,\beta) =  \epsilon Q(u_{(0,0)},u_{(0,1)},u_{(1,0)},u_{(1,1)};\beta,\alpha) \\
 \phantom{Q(u_{(0,0)},u_{(1,0)},u_{(0,1)},u_{(1,1)};\alpha,\beta)} =  \sigma Q(u_{(1,0)},u_{(0,0)},u_{(1,1)},u_{(0,1)};\alpha,\beta) \,,
\end{eqnarray*}
where $\epsilon = \pm 1$ and $\sigma = \pm 1$.
\end{enumerate}

\begin{figure}[h]
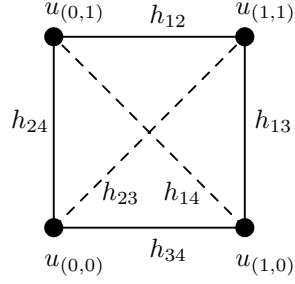

\centertexdraw{ \setunitscale 0.5
\linewd 0.02 \arrowheadtype t:F 
\htext(0 0.5) {\phantom{T}}
\move (-1 -2) \lvec (1 -2) 
\move(-1 -2) \lvec (-1 0) \move(1 -2) \lvec (1 0) \move(-1 0) \lvec(1 0)
\move (1 -2) \fcir f:0.0 r:0.1 \move (-1 -2) \fcir f:0.0 r:0.1
 \move (-1 0) \fcir f:0.0 r:0.1 \move (1 0) \fcir f:0.0 r:0.1  
 \move (-1 -2) \lpatt (.1 .1) \lvec (1 0) \move (1 -2) \lvec (-1 0) \lpatt ()
\htext (-1.1 -2.5) {$u_{(0,0)}$} \htext (.9 -2.5) {$u_{(1,0)}$} \htext (0 -2.35) {$h_{34}$}
\htext (-1.1 .15) {$u_{(0,1)}$} \htext (.9 .15) {$u_{(1,1)}$} \htext (0 .1) {$h_{12}$}
\htext (-1.45 -1) {$h_{24}$} \htext (1.1 -1) {$h_{13}$}
\htext (-.5 -1.75) {$h_{23}$} \htext (.15 -1.75 ) {$h_{14}$}}
\caption{{\em{The elementary quadrilateral and the polynomials}}} \label{polyhg}
\end{figure}

The affine linearity of $Q$ implies that one can define six different polynomials in terms of the function $Q$ \cite{ABS,ABS1,TTX}, four of them assigned to the edges and the rest to the diagonals of the elementary quadrilateral where the equation is defined, see Figure \ref{polyhg}. 

A polynomial $h_{i j}$ assigned to an edge or a diagonal depends on the values of $u$ assigned to the end-points of the corresponding edge or diagonal, respectively, as illustrated in Figure \ref{polyhg}, and is defined by
$$h_{ij}\,=\,h_{j\,i}\,:=\,Q\,Q_{,i j}\,-\,Q_{,i}\,Q_{,j}\,,\quad i\,\ne\,j\,,\quad i,\,j\,=\,1,\ldots,\,4,$$
where $Q_{,i}$ denotes the derivative of $Q$ with respect to its $i$-th argument and $Q_{,i j}$ the second order derivative $Q$ with respect to its $i$-th and $j$-th argument. The polynomials $h_{i j}$ are quadratic in each one of their arguments. Moreover, the relations
\be h_{12}\,h_{34}\,=\,h_{13}\,h_{24}\,=\,h_{14}\,h_{23} \label{relhG} \ee
hold in view of the condition $Q\,=\,0$.

On the other hand, the symmetries of the square imply the following: 
\begin{enumerate}[1)]
\item The polynomials on the edges have to be of the form
\be
 h_{ij}\,=\,\left\{\begin{array}{l c} h(x,y;\alpha,\beta), & |i-j|\,=\,1 \\ h(x,y;\beta,\alpha), & |i-j|\,=\,2 \end{array} \right.\,,\quad i \ne j \,,\quad \{i,j\} \ne \{2,3\}\,, \label{hijedgesh}
\ee
where $h$ is quadratic and symmetric in its first two arguments.
\item The two diagonal polynomials are identical and
\be h_{14}\,=\,h_{23}\,=\, G(x,y;\alpha,\beta)\,, \label{hijdiagG} \ee
where $G$ is quadratic, symmetric in its first two arguments and symmetric in the parameters.
\end{enumerate}

\subsection{The Adler--Bobenko--Suris equations}

In order to make our presentation self-contained, we first list all the members of the ABS classification and the notation that we will use in the next sections:
\begin{eqnarray} 
&{\mbox{\it{H1}}}& (u_{(0,0)}-u_{(1,1)})\, (u_{(1,0)}-u_{(0,1)})\, -\,\alpha \,+ \, \beta \, = \,0 \label{H1} \\
&&\nonumber  \\
&{\mbox{\it{H2}}}& (u_{(0,0)}-u_{(1,1)})(u_{(1,0)}-u_{(0,1)}) +(\beta-\alpha) (u_{(0,0)}+u_{(1,0)}+u_{(0,1)}+u_{(1,1)})  \nonumber \\
&&\qquad \,\, - \alpha^2 + \beta^2 = 0 \label{H2} \\
&&\nonumber \\
&{\mbox{\it{H3}}}& \alpha (u_{(0,0)} u_{(1,0)}+u_{(0,1)} u_{(1,1)}) - \beta (u_{(0,0)} u_{(0,1)}+u_{(1,0)} u_{(1,1)}) + \delta\, (\alpha^2-\beta^2) \,= \,0\label{H3}
\end{eqnarray} 
\begin{eqnarray} 
&{\mbox{\it{Q1}}}& \alpha (u_{(0,0)}-u_{(0,1)}) (u_{(1,0)}- u_{(1,1)}) - \beta (u_{(0,0)}- u_{(1,0)}) (u_{(0,1)} -u_{(1,1)}) + \delta^2 \alpha \beta (\alpha-\beta)= 0 \label{Q1} \\
&& \nonumber \\
&{\mbox{\it{Q2}}}& \alpha (u_{(0,0)}-u_{(0,1)}) (u_{(1,0)}- u_{(1,1)}) - \beta (u_{(0,0)}- u_{(1,0)}) (u_{(0,1)} -u_{(1,1)}) \nonumber\\
&& + \alpha \beta (\alpha-\beta) (u_{(0,0)}+u_{(1,0)}+u_{(0,1)}+u_{(1,1)}) - \alpha \beta (\alpha-\beta) (\alpha^2-\alpha \beta + \beta^2) = 0 \label{Q2}\\
&& \nonumber \\
&{\mbox{\it{Q3}}}& (\beta^2-\alpha^2) (u_{(0,0)} u_{(1,1)}+u_{(1,0)} u_{(0,1)}) + \beta (\alpha^2-1) (u_{(0,0)} u_{(1,0)}+u_{(0,1)} u_{(1,1)}) \nonumber\\
&& - \alpha (\beta^2-1) (u_{(0,0)} u_{(0,1)}+u_{(1,0)} u_{(1,1)}) - \frac{\delta^2 (\alpha^2-\beta^2) (\alpha^2-1) (\beta^2-1)}{4 \alpha \beta}=0  \label{Q3} \\
&& \nonumber \\
&{\mbox{\it{Q4}}}& a_0 u_{(0,0)} u_{(1,0)} u_{(0,1)} u_{(1,1)} \nonumber \\
&& + a_1 (u_{(0,0)} u_{(1,0)} u_{(0,1)} + u_{(1,0)} u_{(0,1)} u_{(1,1)} + u_{(0,1)} u_{(1,1)} u_{(0,0)} + u_{(1,1)} u_{(0,0)} u_{(1,0)}) \nonumber \\
&& +\alpha_2 (u_{(0,0)} u_{(1,1)} + u_{(1,0)} u_{(0,1)}) + \bar{a}_2 (u_{(0,0)} u_{(1,0)}+u_{(0,1)} u_{(1,1)})  \label{Q4} \\
&& + \tilde{a}_2 (u_{(0,0)} u_{(0,1)}+u_{(1,0)} u_{(1,1)}) + a_3 (u_{(0,0)} + u_{(1,0)} + u_{(0,1)} + u_{(1,1)}) + a_4 = 0\nonumber
\end{eqnarray}
The $a_i$'s appearing in the last equation are determined by the relations
$$ a_0 = a+b \,,\,\,\,a_1=-a \beta - b \alpha\,,\,\,\,a_2=a \beta^2 + b \alpha^2\,,$$
$$ \bar{a}_2 = \frac{a b (a+b)}{2 (\alpha-\beta)} + a \beta^2 - \left(2 \alpha^2 - \frac{g_2}{4}\right) b\,,\,\,
 \tilde{a}_2 = \frac{a b (a+b)}{2 (\beta-\alpha)} + b \alpha^2 - \left(2 \beta^2 - \frac{g_2}{4}\right) a\,, $$
$$ a_3 = \frac{g_3}{2}a_0 - \frac{g_2}{4} a_1\,,\,\,\,a_4=\frac{g_2^2}{16}a_0-g_3 a_1\,,$$
where
$$a^2\, =\, p(\alpha)\,,\quad b^2 \,=\, p(\beta)\,,\quad p(x)\,=\,4 x^3-g_2 x - g_3\,.$$

The main characteristic of all of the above equations is their integrability, which is understood as their being multidimensionally consistent. From this property it follows that \cite{ABS}:
\begin{enumerate}[i)]
\item The polynomial related to the edges, $h$, can be written as
$$h(x,y;\alpha,\beta)\,=\,k(\alpha,\beta) \,f(x,y,\alpha)\,,$$ 
where the function $k(\alpha,\beta)$ is antisymmetric, i.e.
$$k(\beta,\alpha) \,= \, - k(\alpha,\beta)\,.$$
\item The discriminant 
$$d\,:=\, f_{,y}^2 \,-\, 2\, f\, f_{,y y}$$
is independent of the parameters $\alpha$, $\beta$.
\item The functions $f$, $G$ and $k$ determining the polynomials $h_{i j}$ can be specified explicitly and, for convenience, are given in \ref{fGk}.
\end{enumerate}

To the above properties of the ABS equations, one can add the following two \cite{Xen}, which will also be used in the symmetry analysis to be presented in the following sections.

\begin{itemize}[iv)]
\item They define their own auto-B{\"a}cklund transformation. The latter is specified by the following relations
\be
{\mathds{B}}_d(u,\tilde{u},\lambda) \,:=\,\left\{ \begin{array}{l} Q(u_{(0,0)},u_{(1,0)},\tilde{u}_{(0,0)},\tilde{u}_{(1,0)};\alpha,\lambda) \,=\, 0\\
Q(u_{(0,0)},u_{(0,1)},\tilde{u}_{(0,0)},\tilde{u}_{(0,1)};\beta,\lambda) \,=\, 0 \end{array} \right. \,.\label{discautoBac}
\ee
\end{itemize}
\begin{itemize}[v)]
\item If $\{u^0,u^1,u^2,u^{12}\}$ is a quartet of solutions related by the B{\"a}cklund transformation ${\mathds{B}}_d$, then their superposition (Bianchi diagram) is expressed by the condition
$$Q\left(u^0,u^1,u^2,u^{12};\lambda_1,\lambda_2\right)\,=\,0\,.$$
\end{itemize}

\section{Symmetry reductions} \label{symred}

In this section we present the general framework of particular symmetry reductions of the ABS equations. More specifically, we study solutions of these equations which remain invariant under the action of both of the extended three point generalized symmetry generators, under the assumption that the unknown function depends continuously on the lattice parameters $\alpha$, $\beta$. 

We first show that, invariant solutions of the above kind are determined by a system of differential--difference equations. The latter turns out to be equivalent to an integrable system of partial differential equations, $\Sigma[u]$. The integrability of $\Sigma[u]$ is established by the construction of its auto-B{\"a}cklund transformation, ${\mathds{B}}_c$. This transformation provides the means for deriving a Lax pair for system $\Sigma[u]$, as well. These integrability aspects are the subject of the second part of this section.

\subsection{Continuous symmetry reductions and system $\Sigma[u]$ : general considerations}

Let us recall that \cite{TTX}, every integrable lattice equation
\be Q(u_{(0,0)},u_{(1,0)},u_{(0,1)},u_{(1,1)};\alpha,\beta)\,=\,0 \label{HQinvsol} \ee
admits a pair of three point generalized symmetries generated, respectively, by the vector fields 
\begin{subeqnarray} \label{genvnvm1}
{\bf v}_n &=& R(u_{(0,0)},u_{(1,0)},u_{(-1,0)},\alpha)\,\partial_{u_{(0,0)}} \,, \label{genvn11} \\
{\bf v}_m &=& R(u_{(0,0)},u_{(0,1)},u_{(0,-1)},\beta)\,\partial_{u_{(0,0)}}\,. \label{genvm12} 
\end{subeqnarray}
It also admits a pair of extended generalized symmetries with respective generators the vector fields
\begin{subeqnarray} \label{extgenv1v21}
 {\mathbf{v}}_1 \,=\, A(n)\,R(u_{(0,0)},u_{(1,0)},u_{(-1,0)},\alpha)\,\partial_{u_{(0,0)}} \,+\, (A(n)-A(n+1))\, r(\alpha)\, \partial_\alpha\,,\label{extgenv11} \\
 {\mathbf{v}}_2 \,=\,  B(m)\,R(u_{(0,0)},u_{(0,1)},u_{(0,-1)},\beta)\,\partial_{u_{(0,0)}}\,+\,(B(m)-B(m+1))\, r(\beta)\, \partial_\beta\,, \label{extgenv21}
\end{subeqnarray}
where
\be  R(u,x,y,\kappa)\,:=\, \frac{f(u,x,\kappa)}{x-y}\,-\,\frac{1}{2}f_{,x}(u,x,\kappa)\,=\,\frac{f(u,y,\kappa)}{x-y}\,+\,\frac{1}{2}f_{,y}(u,y,\kappa)\,, \label{genR} \ee
$A(n)$, $B(m)$ are arbitrary non-constant functions of their arguments, and $r(x)$ depends on the particular equation under consideration, as specified in the following table:\vspace{.2cm}

\begin{center}
\begin{tabular}{|c|c|c|c|c|c|c|c|}
\hline Equation & {\it H1} & {\it H2} & {\it H3} & {\it Q1} & {\it Q2} & {\it Q3} & {\it Q4}  \\ \hline
 $r(x)$ & $1$ & $1$ & $-\,\frac{x}{2}$ & $1$ & $1$ &  $-\,\frac{x}{2}$ & $-\frac{1}{2}\,(4 x^3-g_2 x-g_3)^{1/2}$ \\ \hline
\end{tabular}
\end{center}
\vspace{.2cm}

The solutions of (\ref{HQinvsol}) that remain invariant under the action of both of the symmetry generators ${\mathbf{v}}_1$ and ${\mathbf{v}}_2$ must satisfy the invariant surface conditions
\begin{subequations} \label{parcon}
 \begin{eqnarray}
\frac{\partial\,u_{(0,0)}}{\partial\,\alpha}&=& K(n,\alpha)\,R(u_{(0,0)},u_{(1,0)},u_{(-1,0)},\alpha)\,, \label{parcon1} \\
\frac{\partial\,u_{(0,0)}}{\partial\,\beta} &=& L(m,\beta)\,R(u_{(0,0)},u_{(0,1)},u_{(0,-1)},\beta)\,, \label{parcon2}
 \end{eqnarray}
\end{subequations}
where
\be\begin{array}{l}
K(n,\alpha)\,:=\,\frac{A(n)}{A(n)-A(n+1)}\,\frac{1}{r(\alpha)}\,,\\
L(m,\beta)\,:=\,  \frac{B(m)}{B(m)-B(m+1)}\,\frac{1}{r(\beta)}\,.  \end{array} \label{parconKL}\ee

The compatibility of equations (\ref{HQinvsol}) and (\ref{parcon}) is expressed by the conditions
\begin{subequations} \label{parcomcon}
\begin{eqnarray}
{\rm{D}}_\alpha\left(Q(u_{(0,0)},u_{(1,0)},u_{(0,1)},u_{(1,1)};\alpha,\beta)\right) \,=\,0\,,\label{parcomcon1}\\
{\rm{D}}_\beta\left(Q(u_{(0,0)},u_{(1,0)},u_{(0,1)},u_{(1,1)};\alpha,\beta)\right) \,=\,0\,,\label{parcomcon2}\\
\partial_\beta \left(\partial_\alpha u_{(0,0)} \right) \,=\, \partial_\alpha \left(\partial_\beta u_{(0,0)} \right)\,,\label{parcomcon3}
\end{eqnarray}
\end{subequations}
where ${\rm{D}}_\alpha$ and ${\rm{D}}_\beta$ denote the total derivative operators with respect to $\alpha$ and $\beta$, respectively, i.e.
$${\rm{D}}_\alpha\,:=\,\partial_\alpha\,+\,\sum_{i,j=0}^{1}\frac{\partial u_{(i,j)}}{\partial \alpha} \partial_{u_{(i,j)}}\,,\quad {\rm{D}}_\beta\,:=\,\partial_\beta\,+\,\sum_{i,j=0}^{1}\frac{\partial u_{(i,j)}}{\partial \beta} \partial_{u_{(i,j)}}\,. $$

Written out explicitly, by using the expressions for $Q_{,\alpha}$ and $Q_{,\beta}$ following from the determining equations for the symmetry generators ${\bf v}_1$ and ${\bf v}_2$, respectively, conditions (\ref{parcomcon1}), (\ref{parcomcon2}) imply that $A(n)$ and $B(m)$ must be affine linear. Without loss of generality, we choose them to read as follows:
$$A(n)\,=\,n\,,\quad B(m)\,=\,m\,.$$

Condition (\ref{parcomcon3}), on the other hand, imposes no further restrictions, because it holds identically. This follows from the fact that, the commutator of the two symmetry generators ${\bf v}_n$, ${\bf v}_m$ produces a trivial generalized symmetry \cite{Xen}.

Thus, the solutions of the ABS equations which are invariant under the action of both ${\bf v}_1$ and ${\bf v}_2$ are determined by the differential--difference system 
\begin{subequations} \label{parconf}
 \begin{eqnarray}
\qquad \quad\, Q(u_{(0,0)},u_{(1,0)},u_{(0,1)},u_{(1,1)};\alpha,\beta)&=&0 \,,\label{parconfeq}\\
r(\alpha)\,\frac{\partial\,u_{(0,0)}}{\partial\,\alpha}\,+\,n\,R(u_{(0,0)},u_{(1,0)},u_{(-1,0)},\alpha) &=& 0\,, \label{parconf1} \\
r(\beta)\,\frac{\partial\,u_{(0,0)}}{\partial\,\beta} \,+\,m \,R(u_{(0,0)},u_{(0,1)},u_{(0,-1)},\beta) &=&0\,. \label{parconf2}
 \end{eqnarray}
\end{subequations}
These solutions will be referred to as {\sl{continuously invariant solutions}}.

System (\ref{parconf}) involves the values of the unknown function $u$ at six different points of the lattice. One could eliminate any three of these values and get an equivalent system of {\bf{partial differential equations}} involving the remaining ones. We choose to eliminate the values $u_{(-1,0)}$, $u_{(0,-1)}$ and $u_{(1,1)}$, and this leads to the following result.

\begin{thm}
Every continuous invariant solution is determined by the system of partial differential equations 

$$\frac{\partial u_{(1,0)}}{\partial \beta}=\frac{G(u_{(1,0)},u_{(0,1)})}{k(\alpha,\beta) f(u_{(0,0)},u_{(0,1)},\beta)}\frac{\partial u_{(0,0)}}{\partial \beta} +\frac{m f(u_{(0,0)},u_{(0,1)},\beta)}{2 r(\beta) k(\alpha,\beta)}\partial_{u_{(0,1)}}\left(\frac{G(u_{(1,0)},u_{(0,1)})}{f(u_{(0,0)},u_{(0,1)},\beta)} \right),$$

$$\frac{\partial u_{(0,1)}}{\partial \alpha}=\frac{G(u_{(1,0)},u_{(0,1)})}{k(\beta,\alpha) f(u_{(0,0)},u_{(1,0)},\alpha)}\frac{\partial u_{(0,0)}}{\partial \alpha} +\frac{n f(u_{(0,0)},u_{(1,0)},\alpha)}{2 r(\alpha) k(\beta,\alpha)}\partial_{u_{(1,0)}}\left(\frac{G(u_{(1,0)},u_{(0,1)})}{f(u_{(0,0)},u_{(1,0)},\alpha)} \right),$$

\begin{eqnarray*}
 \frac{\partial^2 u_{(0,0)}}{\partial \alpha \partial \beta} = A_1  \frac{\partial u_{(0,0)}}{\partial \alpha}\frac{\partial u_{(0,0)}}{\partial \beta} + \frac{f(u_{(0,0)},u_{(1,0)},\alpha)}{2 k(\alpha,\beta)}\left( \frac{m}{r(\beta)} A_2 \frac{\partial u_{(0,0)}}{\partial \alpha} + \frac{n}{r(\alpha)} A_3 \frac{\partial u_{(0,0)}}{\partial \beta}\right)\\
\\
 {\phantom{\frac{\partial^2 u_{(0,0)}}{\partial \alpha \partial \beta} =}} + \, \frac{n m f(u_{(0,0)},u_{(1,0)},\alpha)}{4 r(\alpha) r(\beta) k(\alpha,\beta)}A_4\,,\qquad
\end{eqnarray*}
where

\begin{eqnarray*}
 A_1 = \left(\frac{f_{,u_{(0,0)}}(u_{(0,0)},u_{(1,0)},\alpha)}{f(u_{(0,0)},u_{(1,0)},\alpha)} - \frac{f(u_{(0,0)},u_{(1,0)},\alpha)}{k(\alpha,\beta) f(u_{(0,0)},u_{(0,1)},\beta)}\partial_{u_{(1,0)}}\left(\frac{G(u_{(1,0)},u_{(0,1)})}{f(u_{(0,0)},u_{(1,0)},\alpha)} \right) \right)\,,\\
\\
\\
 A_2 = \partial_{u_{(1,0)}} \left( \frac{f_{,u_{(0,1)}}(u_{(0,0)},u_{(0,1)},\beta)}{f(u_{(0,0)},u_{(0,1)},\beta)}\,\frac{G(u_{(1,0)},u_{(0,1)})}{f(u_{(0,0)},u_{(1,0)},\alpha)}  - \frac{G_{,u_{(0,1)}}(u_{(1,0)},u_{(0,1)})}{f(u_{(0,0)},u_{(1,0)},\alpha)} \right)\,,\\
\end{eqnarray*}

\begin{eqnarray*}
 A_3 =  k(\alpha,\beta) \partial_{u_{(1,0)}}\left(\ln f(u_{(0,0)},u_{(1,0)},\alpha)\right) - \frac{f(u_{(0,0)},u_{(1,0)},\alpha)}{f(u_{(0,0)},u_{(0,1)},\beta)} \partial_{u_{(1,0)}}\left(\frac{G_{,u_{(1,0)}}(u_{(1,0)},u_{(0,1)})}{f(u_{(0,0)},u_{(1,0)},\alpha)}\right) \\
\\
 {\phantom{A_3 =}} + \frac{G(u_{(1,0)},u_{(0,1)})}{f(u_{(0,0)},u_{(0,1)},\beta)} \partial^2_{u_{(1,0)}} \left(\ln f(u_{(0,0)},u_{(1,0)},\alpha)\right)\,,
\end{eqnarray*}
and

\begin{eqnarray*}
 A_4 =  \partial_{u_{(1,0)}}  \partial_{u_{(0,1)}}\left(\frac{f_{,u_{(1,0)}}(u_{(0,0)},u_{(1,0)},\alpha) G(u_{(1,0)},u_{(0,1)})}{f(u_{(0,0)},u_{(1,0)},\alpha)}-G_{,u_{(1,0)}}(u_{(1,0)},u_{(0,1)}) \right) \\
 \\
{\phantom{A_3 =}} + \partial_{u_{(0,1)}}\left(\ln f(u_{(0,0)},u_{(0,1)},\beta) \right) f(u_{(0,0)},u_{(1,0)},\alpha) \partial_{u_{(1,0)}} \left( \frac{G_{,u_{(1,0)}}(u_{(1,0)},u_{(0,1)})}{f(u_{(0,0)},u_{(1,0)},\alpha)} \right) \\
\\
{\phantom{A_3 =}} - \partial_{u_{(0,1)}}\left(\ln  f(u_{(0,0)},u_{(0,1)},\beta) \right) 
\partial^2_{u_{(1,0)}}\left(\ln f(u_{(0,0)},u_{(1,0)},\alpha)  \right) G(u_{(1,0)},u_{(0,1)})\,.
\end{eqnarray*}

\hspace{-.8cm}The above system, which will be denoted by $\Sigma\left(u_{(0,0)},u_{(1,0)},u_{(0,1)};\alpha,\beta;n,m\right)$, or, simply $\Sigma[u]$, is symmetric:
$$\Sigma\left(u_{(0,0)},u_{(0,1)},u_{(1,0)};\beta,\alpha;m,n\right)\,=\,\Sigma\left(u_{(0,0)},u_{(1,0)},u_{(0,1)};\alpha,\beta;n,m\right) \,.$$
\end{thm}

\begin{pr}
The first equation of system $\Sigma[u]$ results by eliminating the values $u_{(0,-1)}$ and $u_{(1,-1)}$ from equation (\ref{parconf2}) and
$$Q(u_{(0,-1)},u_{(1,-1)},u_{(0,0)},u_{(1,0)};\alpha,\beta)\,=\,0.$$
Using the affine linearity and the symmetries of $Q$, the last equation can be written as
\be \quad u_{(1,-1)}\,=\,-\,\frac{Q_{,u_{(0,1)}} u_{(0,-1)}+Q}{Q_{,u_{(0,1)} u_{(1,1)}} u_{(0,-1)}+Q_{,u_{(1,1)}}}\,, \label{S1u1t}\ee
where the arguments of $Q(u_{(0,0)},u_{(1,0)},u_{(0,1)},u_{(1,1)};\alpha,\beta)$ have been omitted and $Q$ and its derivatives are understood to be evaluated at $u_{(0,1)}=u_{(1,1)}=0$.

We now solve (\ref{parconf2}) and its shift in the $n$ direction with respect to $u_{(0,-1)}$ and $u_{(1,-1)}$, respectively, and substitute the results into equation (\ref{S1u1t}). The resulting equation, combined with the relations
\begin{eqnarray*}
 Q_{,u_{(1,1)}}^2 &=& \frac{f(u_{(0,0)},u_{(1,0)},\alpha) G(u_{(1,0)},u_{(0,1)})}{f(u_{(1,0)},u_{(1,1)},\beta)}\,,\\
 \partial_{u_{(0,1)}}Q_{,u_{(1,1)}}^2&=& \frac{k(\alpha,\beta)\,f(u_{(0,0)},u_{(1,0)},\alpha)\left(G_{,u_{(0,1)}}(u_{(1,0)},u_{(0,1)})-f_{,u_{(1,1)}}(u_{(1,0)},u_{(1,1)},\beta)\right)}{f(u_{(1,0)},u_{(1,1)},\beta)}\,,
\end{eqnarray*}
which hold in view of $Q=0$, yields the first member of $\Sigma[u]$. 

The second equation of $\Sigma[u]$ results in a similar manner. It is also easily verified that, the first two equations of $\Sigma[u]$ are symmetric, i.e. the one is mapped to the other under interchanges
\be
u_{(1,0)}\,\longleftrightarrow\,u_{(0,1)}\,,\quad \alpha \,\longleftrightarrow \,\beta\,,\quad n\,\longleftrightarrow\,m\,. \label{changesS}
\ee

In order to obtain the third member of $\Sigma[u]$, one first solves the second equation of $\Sigma[u]$ and its shift in the $n$ direction for $\partial_\beta u_{(1,0)}$ and $\partial_\beta u_{(-1,0)}$. One then substitutes the result into the derivative of equation (\ref{parconf1}) with respect to $\beta$. From the resulting equation, one arrives at the third member of $\Sigma[u]$ by using the expressions for $u_{(1,0)}-u_{(-1,0)}$, $G(u_{(-1,0)},u_{(0,1)})$ and its derivatives provided by equation (\ref{parconf1}) and the relation{\footnote{This relation holds identically, i.e. without taking into account the equation $Q=0$, thus we can differentiate it assuming that the corresponding values of $u$ are independent.}}
$$G_{,u_{(1,0)}}(u_{(1,0)},u_{(0,1)})+G_{,u_{(-1,0)}}(u_{(-1,0)},u_{(0,1)})=2\,\frac{G(u_{(1,0)},u_{(0,1)}) - G(u_{(-1,0)},u_{(0,1)})}{u_{(1,0)} - u_{(-1,0)}},$$
and its differential consequences, respectively.

Finally, differentiating equation (\ref{parconf2}) with respect to $\alpha$ and following an analogous procedure, one arrives at an expression which is identical to the third member of $\Sigma[u]$ under the mapping (\ref{changesS}). \hfill $\Box$
\end{pr}

As already noted, one may choose to eliminate any other triad of the values of $u$ involved in equations (\ref{parconf}). In this fashion, compatible systems of partial differential equations can be constructed involving the triplets $(u_{(0,0)}$, $u_{(-1,0)}$, $u_{(0,1)})$, $(u_{(0,0)}$, $u_{(1,0)}$, $u_{(0,-1)})$ and $(u_{(0,0)}$, $u_{(-1,0)}$, $u_{(0,-1)})$, respectively. It turns out that, the corresponding systems are given by
$$\Sigma(u_{(0,0)},u_{(-1,0)},u_{(0,1)};\alpha,\beta;-n,m)\,,\quad \Sigma(u_{(0,0)},u_{(1,0)},u_{(0,-1)};\alpha,\beta;n,-m) $$
and
$$\Sigma(u_{(0,0)},u_{(-1,0)},u_{(0,-1)};\alpha,\beta;-n,-m)\,,$$
respectively.

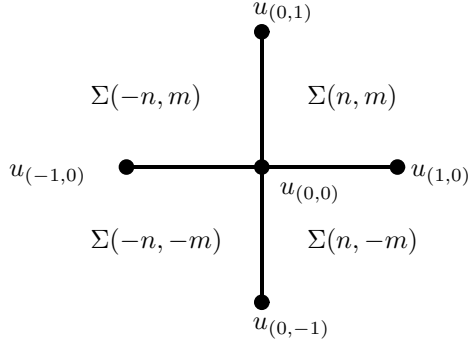
\begin{figure}[h]
\begin{center}
\setlength{\unitlength}{1.2cm}%
\begingroup\makeatletter\ifx\SetFigFont\undefined%
\gdef\SetFigFont#1#2#3#4#5{%
  \reset@font\fontsize{#1}{#2pt}%
  \fontfamily{#3}\fontseries{#4}\fontshape{#5}%
  \selectfont}%
\fi\endgroup%
\begin{picture}(3,3.5)(0,0)
\thicklines
\put(1.5,0){\line( 0,0){3}}
\put(0,1.5){\line(1,0){3}}
\put(1.5,1.5){\circle*{.15}}
\put(1.5,0){\circle*{.15}}
\put(1.5,3){\circle*{.15}}
\put(0,1.5){\circle*{.15}}
\put(3,1.5){\circle*{.15}}
\put(1.7,1.2){\makebox(0,0)[lb]{\smash{\SetFigFont{10}{12}{\rmdefault}{\mddefault}{\updefault}$u_{(0,0)}$}}}
\put(3.15,1.4){\makebox(0,0)[lb]{\smash{\SetFigFont{10}{12}{\rmdefault}{\mddefault}{\updefault}$u_{(1,0)}$}}}
\put(-1.3,1.4){\makebox(0,0)[lb]{\smash{\SetFigFont{10}{12}{\rmdefault}{\mddefault}{\updefault}$u_{(-1,0)}$}}}
\put(1.4,3.2){\makebox(0,0)[lb]{\smash{\SetFigFont{10}{12}{\rmdefault}{\mddefault}{\updefault}$u_{(0,1)}$}}}
\put(1.4,-.3){\makebox(0,0)[lb]{\smash{\SetFigFont{10}{12}{\rmdefault}{\mddefault}{\updefault}$u_{(0,-1)}$}}}
\put(2,2.2){\makebox(0,0)[lb]{\smash{\SetFigFont{10}{12}{\rmdefault}{\mddefault}{\updefault}$\Sigma(n,m)$}}}
\put(2,.6){\makebox(0,0)[lb]{\smash{\SetFigFont{10}{12}{\rmdefault}{\mddefault}{\updefault}$\Sigma(n,-m)$}}}
\put(-.4,2.2){\makebox(0,0)[lb]{\smash{\SetFigFont{10}{12}{\rmdefault}{\mddefault}{\updefault}$\Sigma(-n,m)$}}}
\put(-.4,.6){\makebox(0,0)[lb]{\smash{\SetFigFont{10}{12}{\rmdefault}{\mddefault}{\updefault}$\Sigma(-n,-m)$}}}
\end{picture}
\vspace{.3cm}
\caption{The values of $u$ and the compatible systems $\Sigma$} \label{fig:comsys}
\end{center}
\end{figure}

System $\Sigma(u_{(0,0)},u_{(1,0)},u_{(0,1)};\alpha,\beta;n,m)$ and the last three are compatible, in the following sense, cf. Figure \ref{fig:comsys}. If we eliminate $\partial_\beta u_{(1,0)}$ (respectively $\partial_\beta u_{(-1,0)}$) from systems $\Sigma(n,m)$ and $\Sigma(n,-m)$ (respectively $\Sigma(-n,m)$ and $\Sigma(-n,-m)$), then we will end up with (\ref{parconf2}). On the other hand, the elimination of $\partial_\alpha u_{(0,1)}$ and $\partial_\alpha u_{(0,-1)}$ from systems $\Sigma(n,m)$, $\Sigma(-n,m)$ and $\Sigma(n,-m)$, $\Sigma(-n,-m)$, respectively, leads to (\ref{parconf1}). Finally, the elimination of $\partial_\alpha \partial_\beta u_{(0,0)}$ from any two of the four $\Sigma[u]$'s mentioned above results in (\ref{parconfeq}).

\subsection{Integrability of system $\Sigma[u]$}

We have already characterized system $\Sigma[u]$ as integrable. To support this characterization, in the present subsection, we construct an auto-B{\"a}cklund transformation and a Lax pair for the above system.

To this end, let it first be noted that the fact that every integrable lattice equation
\be Q(u_{(0,0)},u_{(1,0)},u_{(0,1)},u_{(1,1)};\alpha,\beta)\,=\,0 \label{autoBacdiseq1} \ee
admits generalized symmetries and extended generalized symmetries with generators the vector fields given in (\ref{genvnvm1}) and (\ref{extgenv1v21}) respectively, has the following consequences.
\begin{enumerate}[1.]
\item The vector fields 
$$ \tilde{\mathbf{v}}_n \,=\, R\left(u_{(0,0)},u_{(1,0)},u_{(-1,0)},\alpha \right)\,\partial_{u_{(0,0)}}\, + \, R\left(\tilde{u}_{(0,0)},\tilde{u}_{(1,0)},\tilde{u}_{(-1,0)},\alpha \right)\,\partial_{\tilde{u}_{(0,0)}}$$
and
$$ \tilde{\mathbf{v}}_1 \,=\, A(n)\,\tilde{\mathbf{v}}_n\,+\,(A(n)-A(n+1))\,r(\alpha)\,\partial_{\alpha}\,,$$
are symmetry generators of the first of the equations making up the auto-B{\"a}cklund transformation ${\mathds{B}}_d$, i.e. of $Q(u_{(0,0)},u_{(1,0)},\tilde{u}_{(0,0)},\tilde{u}_{(1,0)};\alpha,\lambda)=0$. Therefore, relations
$$\begin{array}{l}
\tilde{\mathbf{v}}_n^{(1)}\left(Q(u_{(0,0)},u_{(1,0)},\tilde{u}_{(0,0)},\tilde{u}_{(1,0)};\alpha,\lambda) \right)\,=\,0\,, \\ \tilde{\mathbf{v}}_1^{(1)}\left(Q(u_{(0,0)},u_{(1,0)},\tilde{u}_{(0,0)},\tilde{u}_{(1,0)};\alpha,\lambda) \right)\,=\,0 \end{array}$$
hold in view of $Q(u_{(0,0)},u_{(1,0)},\tilde{u}_{(0,0)},\tilde{u}_{(1,0)};\alpha,\lambda)=0$.
\item The vector fields
\begin{eqnarray*}
\tilde{\mathbf{v}}_m &=& R\left(u_{(0,0)},u_{(0,1)},u_{(0,-1)},\beta \right)\,\partial_{u_{(0,0)}}\, + \, R\left(\tilde{u}_{(0,0)},\tilde{u}_{(0,1)},\tilde{u}_{(0,-1)},\beta \right)\,\partial_{\tilde{u}_{(0,0)}}\,,\\
\tilde{\mathbf{v}}_2 &=& B(m)\,{\mathbf{v}}_m\,+\,(B(m)-B(m+1))\,r(\beta)\,\partial_{\beta}\,,
\end{eqnarray*}
are symmetry generators of the second of the equations of the auto-B{\"a}cklund transformation. As a result, the pair of relations
$$\begin{array}{l}
\tilde{\mathbf{v}}_m^{(1)}\left(Q(u_{(0,0)},u_{(0,1)},\tilde{u}_{(0,0)},\tilde{u}_{(0,1)};\beta,\lambda) \right)\,=\,0\,, \\ \tilde{\mathbf{v}}_2^{(1)}\left(Q(u_{(0,0)},u_{(0,1)},\tilde{u}_{(0,0)},\tilde{u}_{(0,1)};\beta,\lambda) \right)\,=\,0\end{array}$$
hold in view of $Q(u_{(0,0)},u_{(0,1)},\tilde{u}_{(0,0)},\tilde{u}_{(0,1)};\beta,\lambda)=0$.
\end{enumerate}
Using the above observations, one may prove the following proposition.

\begin{thm} \label{propBcB}
The auto-B{\"a}cklund transformation ${\mathds{B}}_d(u,\tilde{u},\lambda)$ maps a continuously invariant solution $u$ to another solution $\tilde{u}$ of the same kind.
\end{thm}
\begin{pr}
It is given in the \ref{proofBcB}. \hfill $\Box$
\end{pr}

An immediate consequence of the this result is described in the following proposition.

\begin{thm} \label{propBTSigma}
\begin{subequations}
If $u$ is a continuously invariant solution, then the system 

\begin{eqnarray}
 Q(u_{(0,0)},u_{(1,0)},\tilde{u}_{(0,0)},\tilde{u}_{(1,0)};\alpha,\lambda) \,=\, 0\,,\\
 Q(u_{(0,0)},u_{(0,1)},\tilde{u}_{(0,0)},\tilde{u}_{(0,1)};\beta,\lambda) \,=\, 0 \,,
\end{eqnarray}

\begin{eqnarray}
 \frac{\partial\,\tilde{u}_{(0,0)}}{\partial\,\alpha} &=& \frac{1}{k(\alpha,\lambda)}  \left(-\,\frac{\partial\,u_{(0,0)}}{\partial\,\alpha}\,+\,\frac{n}{2\,r(\alpha)} f_{,u_{(1,0)}}(u_{(0,0)},u_{(1,0)},\alpha) \right) \frac{G(u_{(1,0)},\tilde{u}_{(0,0)},\alpha,\lambda)}{f(u_{(0,0)},u_{(1,0)},\alpha)} \nonumber \\
 \nonumber \\
 & & -\,\frac{n}{2\,k(\alpha,\lambda)\,r(\alpha)}\,G_{,u_{(1,0)}}(u_{(1,0)},\tilde{u}_{(0,0)},\alpha,\lambda)\,,\label{conBac1}\\
 \nonumber \\
 \nonumber \\
 \frac{\partial\,\tilde{u}_{(0,0)}}{\partial\,\beta} &=& \frac{1}{k(\beta,\lambda)}  \left(-\,\frac{\partial\,u_{(0,0)}}{\partial\,\beta}\,+\,\frac{m}{2\,r(\beta)} f_{,u_{(0,1)}}(u_{(0,0)},u_{(0,1)},\beta) \right) \frac{G(u_{(0,1)},\tilde{u}_{(0,0)},\beta,\lambda)}{f(u_{(0,0)},u_{(0,1)},\beta)} \nonumber \\
 \nonumber \\
 & & -\,\frac{m}{2\,k(\beta,\lambda)\,r(\beta)}\,G_{,u_{(0,1)}}(u_{(0,1)},\tilde{u}_{(0,0)},\beta,\lambda) \,,\label{conBac2}
\end{eqnarray}
\end{subequations}
defines a new solution $\tilde{u}$ of the same kind, and conversely. 

The above system, which will be denoted as ${\mathds{B}}_c(u,\tilde{u},\lambda)$, is symmetric
$${\mathds{B}}_c(\tilde{u},u,\lambda)\,=\,{\mathds{B}}_c(u,\tilde{u},\lambda)\,,$$
and defines an auto-B{\"{a}}cklund transformation of system $\Sigma[u]$. 
\end{thm}

\begin{pr}
It is given in \ref{proofBTSigma}. \hfill $\Box$
\end{pr}

\begin{rem}
\begin{enumerate}[i)]
\item The second pair of equations of system ${\mathds{B}}_c(u,\tilde{u},\lambda)$ follows from the two first equations of $\Sigma[u]$, via the substitutions
$$u_{(1,0)}\,\longrightarrow \, \tilde{u}_{(0,0)}\,,\quad \alpha \,\longrightarrow \,\lambda $$
and
$$u_{(0,1)}\,\longrightarrow \, \tilde{u}_{(0,0)}\,,\quad \beta \,\longrightarrow \,\lambda\,, $$
respectively.

\item The superposition principle of ${\mathds{B}}_d(u,\tilde{u},\lambda)$ implies the corresponding one for ${\mathds{B}}_c(u,\tilde{u},\lambda)$:
\begin{eqnarray*}
Q\left(u^0_{(0,0)},u^1_{(0,0)},u^2_{(0,0)},u^{12}_{(0,0)};\lambda_1,\lambda_2\right) & = & 0\,, \\
Q\left(u^0_{(1,0)},u^1_{(1,0)},u^2_{(1,0)},u^{12}_{(1,0)};\lambda_1,\lambda_2\right) & = & 0\,, \\
Q\left(u^0_{(0,1)},u^1_{(0,1)},u^2_{(0,1)},u^{12}_{(0,1)};\lambda_1,\lambda_2\right) & = & 0\,.
\end{eqnarray*}
 \hfill $\Box$
\end{enumerate}
\end{rem}

Finally, let us consider the following pair of equations
\begin{subeqnarray} \label{laxsigma}
 \Phi_{,\alpha} &=& \frac{1}{k(\alpha,\lambda)}\,
\left( \begin{array}{cc}
-\frac{1}{2}A_{,\tilde{u}_{(0,0)}} & -\frac{1}{2}A_{, \tilde{u}_{(0,0)} \tilde{u}_{(0,0)}} \\
A & \frac{1}{2}A_{,\tilde{u}_{(0,0)}}
\end{array}\right)\,\Phi\,,\\
& & \nonumber \\
& & \nonumber \\
\Phi_{,\beta} &=& \frac{1}{k(\beta,\lambda)}\,\left( \begin{array}{cc}
-\frac{1}{2}B_{,\tilde{u}_{(0,0)}} & -\frac{1}{2}B_{, \tilde{u}_{(0,0)} \tilde{u}_{(0,0)}} \\
B & \frac{1}{2}B_{,\tilde{u}_{(0,0)}}
\end{array}\right)\,\Phi\,,
\end{subeqnarray}
where
\begin{subeqnarray}
 A &:=& \left(-\,\frac{\partial\,u_{(0,0)}}{\partial\,\alpha}\,+\,\frac{n}{2\,r(\alpha)} f_{,u_{(1,0)}}(u_{(0,0)},u_{(1,0)},\alpha) \right)  \frac{G(u_{(1,0)},\tilde{u}_{(0,0)},\alpha,\lambda)}{f(u_{(0,0)},u_{(1,0)},\alpha)} \nonumber\\
 & &  -\,\frac{n}{2\,r(\alpha)}\,G_{,u_{(1,0)}}(u_{(1,0)},\tilde{u}_{(0,0)},\alpha,\lambda)\,, \\
 \nonumber \\
 B &:=& \left(-\,\frac{\partial\,u_{(0,0)}}{\partial\,\beta}\,+\,\frac{m}{2\,r(\beta)} f_{,u_{(0,1)}}(u_{(0,0)},u_{(0,1)},\beta) \right) \frac{G(u_{(0,1)},\tilde{u}_{(0,0)},\beta,\lambda)}{f(u_{(0,0)},u_{(0,1)},\beta)} \nonumber \\
 \nonumber \\
 && -\,\frac{m}{2\,r(\beta)}\,G_{,u_{(0,1)}}(u_{(0,1)},\tilde{u}_{(0,0)},\beta,\lambda) \,,
\end{subeqnarray}
and $A$, $B$ and their derivatives are evaluated at $\tilde{u}_{(0,0)}=0$. Equations (\ref{laxsigma}) constitute a Lax pair for system $\Sigma[u]$. One arrives at this result, essentially, by the inverse of the procedure presented by Crampin in \cite{Crampin}. In any case, it can be easily verified directly, by considering each of the ABS equations, separately.

\section{Continuous invariant solutions of the discrete potential KdV equation}

In the last section, we established the general framework for the special reductions of the ABS equations leading to what we called continuously invariant solutions. In the present section, the above results are applied to equation {\it H1}. The latter will be also referred to as {\it{discrete potential KdV}}, in compliance with the terminology adopted in \cite{NC}, cf. also \cite{Hirota, NQC}.

System $\Sigma[u]$ corresponding to {\it{H1}} is made up of the equations
\begin{subequations} \label{H1sys}
\begin{eqnarray}
\frac{\partial u_1}{\partial \beta} &= & \frac{u_1-u_2}{\alpha-\beta} \, \left(m\,-\,(u_1-u_2)\,\frac{\partial u}{\partial \beta}\right)\,, \label{H1sys1}\\
& & \nonumber\\
\frac{\partial u_2}{\partial \alpha} &= & \frac{u_1-u_2}{\alpha-\beta} \, \left( n \,+\,(u_1-u_2)\,\frac{\partial u}{\partial \alpha} \right)\,, \label{H1sys2} \\
& & \nonumber\\
\frac{\partial^2 u}{\partial \alpha \partial \beta} &=& \frac{1}{\alpha-\beta} \left( 2\,(u_1-u_2) \frac{\partial u}{\partial \alpha} \frac{\partial u}{\partial \beta}\,+\,n \frac{\partial u}{\partial \beta}\,-\,m \frac{\partial u}{\partial \alpha}\right)\,, \label{H1sys3}
\end{eqnarray}
\end{subequations}
where
$$u\,=\,u_{(0,0)}\,, \quad u_1\,=\,u_{(1,0)}\,,\quad u_2\,=\,u_{(0,1)}\,.$$

Obviously, the nonlinear system (\ref{H1sys}) is very hard to solve. However, whole families of solutions can be obtained, in a systematic way, via symmetry analysis. In what follows, we construct multiparameter families of solutions of the above system, using its Lie point symmetries.

System (\ref{H1sys}) admits a five dimensional group of point symmetries generated by the vector fields \cite{TTX2}
$${\bf w}_1 = \partial_\alpha + \partial_\beta\,,\quad {\bf w}_2\,=\,\alpha \partial_\alpha + \beta \partial_\beta + u \partial_u\,, $$
$${\bf w}_3 = \partial_u\,,\quad {\bf w}_4 = \partial_{u_1} + \partial_{u_2}\,,\quad {\bf w}_5\,=\,u \partial_u - u_1 \partial_{u_1} - u_2 \partial_{u_2}\,. $$
It will be shown that, similarity solutions corresponding to above group of symmetries are determined by solutions of the Painlev{\'e} V and VI equations \cite{TTX2}. For easy reference, we note that, the latter equations are given by
\be
 G^{\prime \prime}\, =\, \left(\frac{1}{2 G} + \frac{1}{G-1}\right) {G^{\prime}}^2 \,-\,\frac{1}{y} G^{\prime} \,+\,
\frak{a} \frac{G (G-1)^2}{y^2}\, +\, \frak{b} \frac{(G-1)^2}{y^2 G}\, +\, \frak{c} \frac{G}{y} \,+\,
\frak{d}  \,\frac{G (G+1)}{G-1} \,, \label{PV}
\ee 
and
\begin{eqnarray}
 G^{\prime \prime} \,=\, \frac{1}{2} \left(\frac{1}{G} + \frac{1}{G-1} + \frac{1}{G-y}\right) {G^{\prime}}^2 -
\left(\frac{1}{y} + \frac{1}{y-1} + \frac{1}{G-y}\right) G^{\prime} \nonumber\\
 \label{PVI} \\  
 \phantom{G^{\prime \prime} \,=\, } + \frac{G (G-1) (G-y)}{y^2 (y-1)^2} \left(\frak{a} + \frak{b} \frac{y}{G^2} + \frak{c} \frac{y-1}{(G-1)^2} +
\frak{d} \frac{y (y-1)}{(G-y)^2}\right)\,, \nonumber 
\end{eqnarray}
which will be denoted by ${\cal{P}}_{{\rm V}}(y,G(y);\frak{a},\frak{b},\frak{c},\frak{d})$ and ${\cal{P}}_{{\rm VI}}(y,G(y);\frak{a},\frak{b},\frak{c},\frak{d})$, respectively.

For the same reason, we list here the symmetry generators of the discrete potential KdV equation \cite{TTX,levi-petr1}, which will be used extensively in the rest of this section:
\begin{enumerate}[]
\item $\bullet$ Point symmetries
$${\mathbf{x}}_1 = \partial_{u_{(0,0)}},\,{\mathbf{x}}_2 = (-1)^{n+m} \partial_{u_{(0,0)}},\,{\mathbf{x}}_3 =(-1)^{n+m} u_{(0,0)} \partial_{u_{(0,0)}}\,,$$
$${\mathbf{x}}_4 = u_{(0,0)} \partial_{u_{(0,0)}}  + 2 \alpha \partial_\alpha + 2 \beta \partial_\beta,\,\,{\mathbf{x}}_5 = \partial_\alpha + \partial_\beta\,.$$
\item $\bullet$ Generalized symmetries
 \begin{eqnarray*}
    {\mathbf{v}}_1 &=&  \frac{1}{u_{(1,0)}-u_{(-1,0)}} \partial_{u_{(0,0)}} \,,\,\,\,{\mathbf{v}}_{2}\, =\, n\, {\mathbf{v}}_1 \, +\, \frac{u_{(0,0)}}{2 (\alpha-\beta)}\, \partial_{u_{(0,0)}}\,, \\
    & & \\
    {\mathbf{v}}_3 &=&  \frac{1}{u_{(0,1)}-u_{(0,-1)}} \partial_{u_{(0,0)}} \,,\,\,\,{\mathbf{v}}_{4} =\,m\,{\mathbf{v}}_3 - \frac{u_{(0,0)}}{2 (\alpha-\beta)}\, \partial_{u_{(0,0)}}\,,
    \end{eqnarray*} 
$$ {\mathbf{v}}_5 =  n {\mbox{\bf{v}}}_1 - \partial_\alpha ,\,\,{\mathbf{v}}_6 =  m {\mbox{\bf{v}}}_3 - \partial_\beta. $$ 
\end{enumerate}

\subsection{Solutions related to Painlev{\'e} V}

We first consider solutions of system (\ref{H1sys}) that are invariant under the action of the symmetry generator 
\be  {\bf{w}}_1\,+\,2\,\mu\,{\bf w}_5\,=\,\partial_\alpha\,+\,\partial_\beta\,+\,2\,\mu\,\left( u\,\partial_u\,-\,u_1\,\partial_{u_1}\,-\,u_2\,\partial_{u_2}\right)\,, \quad \mu \,\in\,{\mathds{R}}-\{0\}\,,  \label{symgenPV} \ee
where $\mu$ may depend on $n$, $m$. Such solutions have to satisfy the differential equations 
$$u_{,\alpha} + u_{,\beta}\,=\,2\,\mu\,u\,,\quad {u_1}_{,\alpha} + {u_1}_{,\beta}\,=\,-2\,\mu\,u_1\,,\quad {u_2}_{,\alpha} + {u_2}_{,\beta}\,=\,-2\,\mu\,u_2\,.$$
This implies that, $u$, $u_1$ and $u_2$ are given by
\be \begin{array}{l} u(\alpha,\beta) \,=\, T_{n,m}(y)\,\exp(\mu z) ,\\
 u_1(\alpha,\beta) \,=\, T_{n+1,m}(y)\,\exp(-\mu z),\\
 u_2(\alpha,\beta)=T_{n,m+1}(y)\,\exp(-\mu z) , \end{array}\label{invformPV} \ee
where
$$y\,=\,\alpha-\beta\,,\quad z\,=\,\alpha+\beta\,,$$
and $T_{i,j}(y)$ are arbitrary functions.

Substitution of these forms into (\ref{H1sys}) yields the following system of ordinary differential equations
\begin{eqnarray}
 y\,\left\{T_{n+1,m}^\prime(y)\,+\,\mu\,T_{n+1,m}(y)\right\} &=& -\,\left(\,m\,+\,{\cal{A}}^{-}\,{\cal{B}}\, \right)\,{\cal{B}}\,,\nonumber\\
 y\,\left\{T_{n,m+1}^\prime(y)\,-\,\mu\,T_{n,m+1}(y)\right\} &=& \left(\,n\,+\,{\cal{A}}^{+}\,{\cal{B}}\, \right)\,{\cal{B}}\,,\label{sysH1redP5}\\
 y \,T_{n,m}^{\prime \prime}(y)\, -\,(n+m) \,T_{n,m}^\prime(y) &=& 2\, {\cal{A}}^{+}\,{\cal{A}}^{-}\,{\cal{B}}\,-\,( n - m - \mu\, y) \,\mu\, T_{n,m}(y)\,,\nonumber
\end{eqnarray}
where
$${\cal{A}}^{\pm}\,:=\,T_{n,m}^\prime(y)\,\pm\,\mu\,T_{n,m}(y)\,,\quad {\cal{B}}\,:=\,T_{n+1,m}(y)\,-\,T_{n,m+1}(y) \,,$$
and the prime denotes differentiation with respect to $y$. The analysis of the latter system can be summarized as follows.

Starting from (\ref{sysH1redP5}) and using differentiation and elimination, one arrives at a fourth order ordinary differential equation for $T_{n,m}(y)$, which is omitted because of its length. The order of the latter differential equation is reduced by one, using the quadrature 
\begin{subequations} \label{SquadPV}
\be  \frac{{\rm d} \phantom{y}}{{\rm d} y} \ln \left(T_{n,m}(y)\right)\,=\,\mu\,\frac{1 + G_{n,m}(y)}{1-G_{n,m}(y)}\,.\label{SquadPV1} \ee
Finally, the resulting third order equation can be integrated once to yield
\be {\cal{P}}_{{\rm V}}\left(y,G_{n,m}(y);\frac{n^2}{2},-\,\frac{m^2}{2}, \lambda,-2\,\mu^2\right)\,, \label{SquadPV2} \ee
\end{subequations}
where $\lambda$ is a constant of integration and may depend on the parameters $n$, $m$.

Returning to system (\ref{sysH1redP5}) and using relations (\ref{SquadPV}), one finds the following expressions for the functions $T_{n+1,m}(y)$, $T_{n,m+1}(y)$ :
\begin{subeqnarray} \label{S1S2PV}
 &&T_{n+1,m}(y) \,=\,  \frac{y G_{n,m}^\prime(y)\,+\,n G_{n,m}^2(y)\,-\,\left(2 n + 1 + \kappa -2 \mu y \right) G_{n,m}(y) + n + \kappa +1}{4 \mu (1-G_{n,m}(y)) T_{n,m}(y)}\,,\label{S1PV}\\
  \nonumber \\
 && T_{n,m+1}(y) \,=\, \frac{y G_{n,m}^\prime(y)\,-\,(m+\kappa +1) G_{n,m}^2(y) \,+\, \left( 2 m + 1 + \kappa + 2 \mu y \right) G_{n,m}(y) - m}{4 \mu (1-G_{n,m}(y)) G_{n,m}(y) T_{n,m}(y)}\,, \label{S2PV}
\end{subeqnarray}
where $\kappa = \lambda/(2 \mu)$.

In order to satisfy the discrete potential KdV equation, the functions $u$, $u_1$ and $u_2$, given by (\ref{invformPV}), (\ref{SquadPV}) and (\ref{S1S2PV}), must be related by shifting appropriately $n$ and $m$, i.e.
$$u_1\,=\,{\cal{S}}_n(u)\,,\quad u_2\,=\,{\cal{S}}_m(u)\,.$$
The above conditions imply that parameter $\mu$ must have the form
$$ \mu\,=\,(-1)^{n+m}\,\tau \,,\quad \tau\,\in\,{\mathds{R}}\,,$$
and functions $T_{i,j}(y)$ have to satisfy the following relations:
$$T_{n+1,m}(y)\,=\,{\cal{S}}_n\left(T_{n,m}(y)\right)\,,\quad T_{n,m+1}(y)\,=\,{\cal{S}}_m\left(T_{n,m}(y)\right)\,.$$

The combination of these conditions with equations (\ref{SquadPV}) and (\ref{S1S2PV}) leads to the following restriction on parameter $\lambda$: It must be of the form
$$\lambda \,=\,\rho\,  - \,(-1)^{n+m}\, \tau \,,\quad \rho\,\in\,{\mathds{R}}\,.$$

Recapitulating, we can state that, the discrete potential KdV equation admits continuously invariant solutions of the form
\begin{subequations} \label{solPV}
\be
u_{(0,0)} \,=\,T_{n,m}(\alpha-\beta)\,\exp\left[(-1)^{n+m} \tau \times \left(\alpha+\beta\right) \right]\,,\quad \tau\,\in\,{\mathds{R}}\,,
\ee
where $T_{n,m}(y)$ is given by the quadrature
\be \frac{{\rm d} \phantom{y}}{{\rm d} y} \ln \left(T_{n,m}(y)\right)\,=\,(-1)^{n+m} \tau \,\frac{1 + G_{n,m}(y)}{1-G_{n,m}(y)}\,, \ee
with $G_{n,m}(y)$ being a solution of the Painlev{\'e} equation
\be {\cal{P}}_{{\rm V}}\left(y,G_{n,m}(y);\frac{n^2}{2},-\,\frac{m^2}{2}, \rho - (-1)^{n+m} \tau ,-2\,\tau^2\right)\,,\quad \rho\,\in\,{\mathds{R}}\,.\ee
\end{subequations}

\begin{rem}
The solutions of the discrete potential KdV equation just constructed can also be considered as being derived from what is referred to as the {\it{asymmetric, alternate discrete Painlev{\'e} II equation}}. This observation results from the following considerations. By construction, any solution $u$ of the class derived above satisfies the differential equation
\be 
\partial_\alpha u_{(0,0)}\,+\,\partial_\beta u_{(0,0)}\,=\,2\,\tau\,(-1)^{n+m}\,u_{(0,0)}\,,\label{cisH1a}
\ee 
as well as, the invariant surface conditions (\ref{parconf1}), (\ref{parconf2}), i.e.
\be
 \partial_\alpha u_{(0,0)} \,+\,\frac{n}{u_{(1,0)}-u_{(-1,0)}}\,=\,0\,,\quad \partial_\beta u_{(0,0)} \,+\,\frac{m}{u_{(0,1)}-u_{(0,-1)}}\,=\,0\,. \label{cisH1}
\ee

Elimination of the derivatives of $u_{(0,0)}$ from (\ref{cisH1a}), (\ref{cisH1}) leads to
\be \frac{n}{u_{(1,0)}-u_{(-1,0)}}\,+\, \frac{m}{u_{(0,1)}-u_{(0,-1)}}\, + \,2 \, \tau \,(-1)^{n+m} \, u_{(0,0)} \,=\,0\,. \label{conaadPII} \ee
This, however, is the invariant surface condition for solutions of the discrete potential KdV that remain invariant under the action of the symmetry generator ${\bf{v}}_2\,+\,{\bf{v}}_4\,+\,2 \tau {\bf{x}}_3$. As shown in \cite{TTX}, this class of group invariant solutions are determined by solutions of the asymmetric, alternate discrete Painlev{\'e} II. \hfill $\Box$
\end{rem}

\subsection{Solutions related to Painlev{\'e} VI}

The solutions of system (\ref{H1sys}) which remain invariant under the action of the symmetry generator 
\be {\bf w}_2 + (2 \mu-1) {\bf w}_5=\alpha \partial_\alpha + \beta \partial_\beta + 2 \mu u \partial_u + \left(1-2 \mu\right) \left( u_1 \partial_{u_1} + u_2 \partial_{u_2}\right), \, \mu  \in {\mathds{R}}-\left\{0\right\}, \label{symgenPVI} \ee
where $\mu$ may depend on $n$ and $m$, must satisfy the differential equations
$$\begin{array}{l}
\alpha u_{,\alpha} + \beta u_{,\beta}=2 \mu u\,,\\
\alpha {u_1}_{,\alpha} + \beta {u_1}_{,\beta} = (1 - 2\mu) u_1\,,\\
\alpha {u_2}_{,\alpha} + \beta {u_2}_{,\beta} = (1 - 2\mu) u_2.\end{array} $$
Hence, these invariant solutions must have the form
\be \begin{array}{l}
u(\alpha,\beta) \,= \,S_{n,m}(y) z^\mu\,,\\
u_1(\alpha,\beta) \,=\, S_{n+1,m}(y) z^{1/2-\mu}\, ,\\
u_2(\alpha,\beta) \,=\, S_{n,m+1}(y) z^{1/2-\mu}\,,\end{array}\label{invformPVI} \ee
where
$$y\,=\,\frac{\alpha}{\beta}\,,\,\,\, z\,=\,\alpha\,\beta\,.$$

Substitution of the above expressions into system (\ref{H1sys}) leads to the system of ordinary differential equations 
\begin{eqnarray*}
&&(1-y)\,\left\{\,2\, y\, S_{n+1,m}^\prime(y)\, +\, (2 \mu -1)\, S_{n+1,m}(y)\, \right\} \,=\, 2\,\left\{\,m\,+\,\sqrt{y} \, {\cal{A}}^{-}\,{\cal{B}}\,\right\}\,{\cal{B}}\,,\\
&&(y-1)\,\left\{\,2\, y\, S_{n,m+1}^\prime(y)\, -\, (2 \mu -1)\, S_{n,m+1}(y)\, \right\} \,=\, 2 \left\{\,n\,y\,+\,\sqrt{y}\,{\cal{A}}^{+}\,{\cal{B}}\,\right\}\,{\cal{B}}\,,\\
&& (y-1)  \left\{y^2  S_{n,m}^{\prime \prime}(y) + y S_{n,m}^\prime(y) - \mu^2 S_{n,m}(y)\right\}  =  2  \sqrt{y}   {\cal{A}}^{+} {\cal{A}}^{-} {\cal{B}} + m {\cal{A}}^{+} + n y {\cal{A}}^{-}\,,
\end{eqnarray*}
where
$${\cal{A}}^{\pm}\,:=\,y\,S_{n,m}^\prime(y)\,\pm\,\mu\,S_{n,m}(y)\,,\quad {\cal{B}}\,:=\,S_{n+1,m}(y)\,-\,S_{n,m+1}(y) \,,$$
and the prime denotes differentiation with respect to $y$. The analysis of the latter system is similar to the one described in the previous subsection regarding system (\ref{sysH1redP5}) and leads to the following results.

The function $S_{n,m}(y)$ is determined by
\begin{subequations} \label{SquadPVI}
\be  \frac{{\rm d} \phantom{y}}{{\rm d} y} \ln \left(S_{n,m}(y)\right)\,=\,\frac{\mu}{y}\,\frac{y + H_{n,m}(y)}{y-H_{n,m}(y)}\,, \label{SquadPVI1} \ee
where the function $H_{n,m}(y)$ is a solution of the equation
\be {\cal{P}}_{{\rm VI}}\left(y,H_{n,m}(y);\frac{n^2}{2},-\,\frac{m^2}{2}, \lambda,\frac{1-4\,\mu^2}{2}\right)\,. \label{SquadPVI2}\ee
\end{subequations}
In the latter, $\lambda$ stands for a constant of integration which may depend on the parameters $n$ and $m$.

The functions $S_{n+1,m}(y)$, $S_{n,m+1}(y)$ are given by
\begin{subeqnarray} \label{S1S2PVI}
&& S_{n+1,m}(y)=  \frac{y^2 (y-1)^2 {H_{n,m}^{\prime}}^2 + 2 (2 \mu -1) y (y-1) H_{n,m} (H_{n,m}-1) H_{n,m}^{\prime} + A_i H_{n,m}^i}{8\,\mu\, (2 \mu -1)\, y^{1/2}\, (H_{n,m} - y)\, (H_{n,m}-1)\, H_{n,m}\, S_{n,m}},  \nonumber \\
\nonumber \\
&& S_{n,m+1}(y) =  \frac{y^2 (y-1)^2 {H_{n,m}^{\prime}}^2 + 2 (2\mu -1) y^2 (y-1) (H_{n,m}-1) H_{n,m}^{\prime} + B_i H_{n,m}^i}{8\,\mu\, (2 \mu -1)\, y^{1/2}\, (H_{n,m}-y)\, (H_{n,m}-1)\, H_{n,m}\, S_{n,m}} ,\nonumber     
\end{subeqnarray}
where we have omitted the argument $y$ of $H_{n,m}$ and $S_{n,m}$. In these relations, summation over the repeated index $i=0,\ldots,4$ is understood and the coefficients $A_i = A_i(y,n,m)$, $B_i = B_i(y,n,m)$ are given by
\begin{eqnarray*}
A_0 (y,n,m) &:=& -m^2 y^2 \,, \\
A_1 (y,n,m)&:=& y \left[\left(m^2  - 2 \lambda + (n -2\mu +1)^2\right) y + 2 m^2 \right] \,, \\  
A_2 (y,n,m)&:=& -(n-2 \mu +1)^2 y^2 - 2 \left(m^2 - 2 \lambda + (n -2\mu +1)^2 \right) y  + (1-2\mu)^2 - m^2,\\
A_3 (y,n,m) &:=& 2 (n-2\mu + 1 )^2 y + m^2 + 2 n^2 -2 \lambda - (n+2 \mu -1)^2  \,,\\
A_4 (y,n,m) &:=& -n (n - 4 \mu +2) \,, \\
B_i (y,n,m) &:=& y^2 A_{4-i} (y^{-1},m,n) \,, \quad i=0, \ldots, 4 \,\,.
\end{eqnarray*} 

\begin{rem} \label{S1S2coneP6}
For later purposes, it is noted that, the functions $S_{n+1,m}(y)$, $S_{n,m+1}(y)$ may also be considered as being determined by the Painlev{\'e} VI transcendent. Specifically, these functions are determined through the quadratures
\begin{subequations} \label{S1S2PVIall}
\be\begin{array}{l}
 \frac{{\rm d} \phantom{y}}{{\rm d} y} \ln \left(S_{n+1,m}(y)\right) \,=\, \frac{1-2 \mu}{2 y}\,\frac{y + H_{n+1,m}(y)}{y -H_{n+1,m}(y)}\,,\\
 \\ 
\frac{{\rm d} \phantom{y}}{{\rm d} y} \ln \left(S_{n,m+1}(y)\right) \,=\, \frac{1 - 2 \mu}{2 y}\,\frac{y + H_{n,m+1}(y)}{y - H_{n,m+1}(y)}\,,\end{array}
\label{S1S2quadPVI} \ee
where $H_{n+1,m}(y)$, $H_{n,m+1}(y)$ satisfy the equations
\be
{\cal{P}}_{{\rm VI}}\left(y,H_{n+1,m}(y);\frac{(n+1)^2}{2},-\,\frac{m^2}{2},\lambda,2\,\mu\,(1-\mu)\right)\,, \label{PVIH1G1G2a}\ee
and
\be 
{\cal{P}}_{{\rm VI}}\left(y,H_{n,m+1}(y);\frac{n^2}{2},-\,\frac{(m+1)^2}{2},\lambda, 2\,\mu\,(1-\mu)\right)\,, \label{PVIH1G1G2b}\ee
respectively. 

This can be proven in the following fashion. Combining  relations (\ref{S1S2PVI}) and (\ref{S1S2quadPVI}), we express $H_{n+1,m}(y)$, $H_{n,m+1}(y)$ in terms of $S_{n,m}(y)$, $G_{n,m}(y)$ and their derivatives.  Substituting the resulting expressions into equations (\ref{PVIH1G1G2a}), (\ref{PVIH1G1G2b}), we arrive at (\ref{SquadPVI2}). \hfill $\Box$
\end{subequations}
\end{rem}
 
In order to satisfy equation {\it H1}, the functions $u$, $u_1$ and $u_2$, determined by (\ref{invformPVI}), (\ref{SquadPVI}) and (\ref{S1S2PVI}), must be such that, $u_1$ and $u_2$ result from $u$ by applying the shift operators on $u$, i.e.
$$u_1\,=\,{\cal{S}}_n(u)\,,\quad u_2\,=\,{\cal{S}}_m(u)\,.$$
These conditions imply that parameter $\mu$ must have the form
$$\mu \,=\,\frac{1}{4}\,\left( 1 \,+\,2 \,(-1)^{n+m} \tau \right)\,,\quad \tau\,\in\,{\mathds{R}}\,,$$
and functions $S_{i,j}(y)$ must satisfy the following relations:
$$S_{n+1,m}(y)\,=\,{\cal{S}}_n\left(S_{n,m}(y)\right)\,,\quad S_{n,m+1}(y)\,=\,{\cal{S}}_m\left(S_{n,m}(y)\right)\,.$$

It is easily verified that, the only consequence of the last conditions is that the parameter $\lambda$ is independent of $n$, $m$. Hence, we conclude that, {\it H1} admits continuously invariant solutions of the form
\begin{subequations}\label{solPVI}
\be
u_{(0,0)}\,=\,S_{n,m}\left(\frac{\alpha}{\beta}\right)\,\left(\alpha \beta \right)^{\left( 1 \,+\,2 \,(-1)^{n+m} \tau \right)/4}\,,\quad \tau \,\in\,{\mathds{R}}\,,\ee
where the function $S_{n,m}(y)$ is determined by the quadrature 
\be \frac{{\rm d} \phantom{y}}{{\rm d} y} \ln \left(S_{n,m}(y)\right)\,=\,\frac{1 \,+\,2 \,(-1)^{n+m} \tau}{4 y}\,\frac{y + H_{n,m}(y)}{y-H_{n,m}(y)}\,, \ee
with $H_{n,m}(y)$ a solution of
\be {\cal{P}}_{{\rm VI}}\left(y,H_{n,m}(y);\frac{n^2}{2},-\,\frac{m^2}{2}, \lambda , \frac{1}{2} - \frac{1}{8}\,\left(1 \,+ \,2\,(-1)^{n+m} \tau \right)^2 \right)\,, \quad \lambda \in {\mathds{R}}.\ee
\end{subequations}

\begin{rem}
Function $u$, defined in (\ref{solPVI}), satisfies the invariant surface conditions (\ref{cisH1}) and, by construction, the differential equation
\be
\alpha\,\partial_\alpha u_{(0,0)}\,+\,\beta\,\partial_\beta u_{(0,0)}\,=\,\left(\frac{1}{2} \,+  \,(-1)^{n+m} \tau \right)\,u_{(0,0)}\,. \label{cisH1P6} 
\ee
Elimination of the derivatives of $u_{(0,0)}$ involved in (\ref{cisH1}) and (\ref{cisH1P6}) leads to
$$\frac{\alpha\, n}{u_{(1,0)}-u_{(-1,0)}}\,+\,\frac{\beta \,m}{u_{(0,1)}-u_{(0,-1)}}\,+\,\left( \frac{1}{2}\,+\,(-1)^{n+m}\tau \right) u_{(0,0)}\,=\,0\,.$$
The last equation implies that, solution (\ref{solPVI}) is also invariant under the action of the generalized symmetry generator $\alpha {\bf{v}}_2 + \beta {\bf{v}}_4 + \tau {\bf{x}}_3$.

Reductions of {\it{H1}} using the above symmetry generator were studied in \cite{TTX}, while the connection of the corresponding similarity solutions to discrete generalized and continuous sixth Painlev{\'e} equations was demonstrated in \cite{NijP6}. On the other hand, similarity solutions corresponding to the symmetry generated by $\alpha {\bf{v}}_2 + \beta {\bf{v}}_4$ were studied in \cite{Pap2}. The latter are contained in the class of solutions given by (\ref{solPVI}) for $\tau  = 0$. \hfill $\Box$
\end{rem}

\section{Continuous invariant solutions of the discrete Schwarzian KdV equation}

The ${\it Q1}_{\delta=0}$ ABS equation, which is also referred to as the {\it{discrete Schwarzian KdV}} \cite{NC}, is given by
\be  \alpha (v_{(0,0)}-v_{(0,1)}) (v_{(1,0)}- v_{(1,1)}) - \beta (v_{(0,0)}- v_{(1,0)}) (v_{(0,1)} -v_{(1,1)}) \,= \,0\,. \label{skdv} \ee
In this section we present continuously invariant solutions of the above equation using similarity solutions of the corresponding system $\Sigma[v]$.

For this reason, we first write out $\Sigma[v]$ explicitly and list the algebra of its Lie point symmetries. Specifically, $\Sigma[v]$ is made up of the equations
\begin{subequations} \label{Q1d0sysf}
\begin{eqnarray}
  \frac{\partial v_1}{\partial \beta} &=& \frac{v_1-v_2}{\alpha-\beta} \, \frac{m\,(v-v_1)\,(v-v_2)\,-\,\beta\,(v_1-v_2)\,\frac{\partial v}{\partial \beta}}{(v-v_2)^2}\,, \\
 \nonumber\\
 \frac{\partial v_2}{\partial \alpha} &=& \frac{v_1-v_2}{\alpha-\beta} \, \frac{n \,(v-v_1)\,(v-v_2)\,+\,\alpha\,(v_1-v_2)\,\frac{\partial v}{\partial \alpha}}{(v-v_1)^2}\,, \\
 \nonumber\\
 \frac{\partial^2 v}{\partial \alpha \partial \beta} &=& \frac{1}{\alpha-\beta} \left( 2\,\left(\frac{\alpha}{v-v_1}\,-\,\frac{\beta}{v-v_2}\right) \frac{\partial v}{\partial \alpha} \frac{\partial v}{\partial \beta}\,+\,n \frac{\partial v}{\partial \beta}\,-\,m \frac{\partial v}{\partial \alpha}\right),
\end{eqnarray}
\end{subequations}
where
$$v\,=\,v_{(0,0)}\,, \quad v_1\,=\,v_{(1,0)}\,,\quad v_2\,=\,v_{(0,1)}\,.$$
Its algebra of Lie point symmetries is three-dimensional and is spanned by the vector fields
$${\bf z}_1\,=\,\alpha \partial_\alpha + \beta \partial_\beta\,,\quad {\bf z}_2\,=\,\partial_v\,+\,\partial_{v_1}\,+\,\partial_{v_2}\,,\quad {\bf z}_3 \,=\,v\, \partial_v\,+\,v_1\,\partial_{v_1}\,+\,v_2\,\partial_{v_2}\,.$$

The solutions of system (\ref{Q1d0sysf}) remaining invariant under the action of the symmetry generator ${\bf z}_1 + 2 \gamma {\bf z}_3$ are determined by the Painlev{\'e} sixth transcendent. To derive this result, we will make use of the following lemma, whose proof is straightforward.
\begin{lm} \label{lemmaH1Q1}
The contact transformation
\begin{subeqnarray} \label{transH1Q1d0}
\frac{\partial v}{\partial \alpha} &=& \frac{u_1 - v}{\alpha}\,\left(n\,+\,(u_1-v)\,\frac{\partial u}{\partial \alpha} \right)\,, \\
\frac{\partial v}{\partial \beta} &=& \frac{u_2 - v}{\beta}\,\left(m\,+\,(u_2-v)\,\frac{\partial u}{\partial \beta} \right)\,, \\
v_1 &=& u_1 \,,\quad v_2 \,=\, u_2\,,
\end{subeqnarray}
maps solutions of system (\ref{H1sys}) to solutions of system (\ref{Q1d0sysf}), and conversely.
\end{lm}

Thus, we start with the similarity solution of system (\ref{H1sys}) specified by (\ref{invformPVI}), (\ref{SquadPVI}) and (\ref{S1S2PVI}), in which we set $\lambda = 2 \ell^2$, for later convenience. Substitution of this solution to (\ref{transH1Q1d0}) and integration of the result leads to the following solution of $\Sigma[v]$:
\begin{eqnarray}
v(\alpha,\beta) &=& P_{n,m}(y)\,z^{-\mu + 1/2}\,,\nonumber \\
v_1(\alpha,\beta) &=& P_{n+1,m}(y)\,z^{-\mu + 1/2},\,\\
v_2(\alpha,\beta) &=& P_{n,m+1}(y)\,z^{-\mu + 1/2},\nonumber
\end{eqnarray}
where $y = \alpha/\beta$, $z = \alpha \beta$, and
\begin{subeqnarray} \label{defStildeall}
&& P_{n,m}(y) \,=\, \frac{(m+n-2 (\ell+\mu)+1) (H_{n,m}(y)-y)}{4 \mu \sqrt{y} (H_{n,m}(y)-1) S_{n,m}(y)}  
 - \, \frac{S_{n+1,m}(y) - H_{n,m}(y) S_{n,m+1}(y)}{H_{n,m}(y)-1}\,,\label{defStilde} \\
&& P_{n+1,m}(y) \,=\, S_{n+1,m}(y)\,, \label{defS1tilde} \\
&& P_{n,m+1}(y) \,=\, S_{n,m+1}(y)\,. \label{defS2tilde}
\end{subeqnarray}

The functions $P_{i,j}(y)$ are determined by solutions of the sixth Painlev{\'e} equation. Regarding $P_{n+1,m}(y)$ and $P_{n,m+1}(y)$, this property is an obvious consequence of equations (\ref{defStildeall}). 

On the other hand, $P_{n,m}(y)$ is determined by the solution $\tilde{H}_{n,m}(y)$ of the Painlev{\'e} VI equation. More specifically, the former is determined by the latter through the quadrature
\begin{subequations}
\be
\frac{{\rm d} \phantom{y}}{{\rm d} y} \ln \left(P_{n,m}(y)\right)\,=\,\frac{1-2 \mu}{2 y}\,\frac{y-\tilde{H}_{n,m}(y)}{y+\tilde{H}_{n,m}(y)}\,,
\ee
where $\tilde{H}_{n,m}(y)$ stands for any solution of the Painlev{\'e} equation
\be
{\cal{P}}_{\rm VI}\left(y,{\tilde{H}}_{n,m}(y);\frac{n^2}{2},-\frac{m^2}{2},\frac{(2 \ell -1)^2}{2},2 \mu (1-\mu) \right)\,. \label{PVIQ1S}
\ee
\end{subequations}
The proof is similar to the one described in Remark \ref{S1S2coneP6}.

So far, we have shown that, the triad of functions
$$\begin{array}{l}
v(\alpha,\beta)\,=\,P_{n,m}(y)\,z^{\gamma}\,,\\
v_1(\alpha,\beta)\,=\,P_{n+1,m}(y)\,z^{\gamma}\,, \\
v_2(\alpha,\beta)\,=\,P_{n,m+1}(y)\,z^{\gamma}\,, \end{array}$$
where
$$\gamma\,=\,-\mu\,+\,\frac{1}{2}\,,$$
and the $P_{i,j}$'s are determined by solutions of the Painlev{\'e} VI equation, provide a solution of system (\ref{Q1d0sysf}). Obviously, all members of this triad are also invariant under the symmetry generator ${\bf z}_1 + 2 \gamma {\bf z}_3$, i.e. they satisfy the differential equations
$$\alpha\,v_{,\alpha}\,+\,\beta\,v_{,\beta}\,=\,2\,\gamma\,v\,,\quad \alpha\,{v_1}_{,\alpha}\,+\,\beta\,{v_1}_{,\beta}\,=\,2\,\gamma\,v_1\,,\quad \alpha\,{v_2}_{,\alpha}\,+\,\beta\,{v_2}_{,\beta}\,=\,2\,\gamma\,v_2\,,$$
respectively.

The functions $v$, $v_1$ and $v_2$, as defined above, form a solution of the discrete Schwarzian KdV equation provided that
$$v_1\,=\,{\cal{S}}_n(v)\,,\quad v_2\,=\,{\cal{S}}_m(v)\,.$$
These conditions imply that, parameter $\gamma$ is independent of $n$, $m$, and parameter $\ell$ must have the form
$$ \ell \,=\,\frac{1}{2}\,\left(\,c \,+\,n\,+\,m\,+\,1\,\right)\,,\quad c \,\in\,{\mathds{R}}\,,$$
in view of which, functions $P_{i,j}(y)$ satisfy the following relations:
$$P_{n+1,m}(y)\,=\,{\cal{S}}_n\left(P_{n,m}(y)\right)\,,\quad P_{n,m+1}(y)\,=\,{\cal{S}}_m\left(P_{n,m}(y)\right)\,.$$

As a result, the continuously invariant solutions of ${\it Q1}_{\delta=0}$ constructed above can be written as
\begin{subequations} \label{invsolQ1P6}
\be v_{(0,0)}\,=\,P_{n,m}\left(\frac{\alpha}{\beta}\right)\,\left(\alpha\,\beta\right)^{\gamma}\,,\quad \gamma \,\in\,{\mathds{R}}\,,\ee
where
\be \frac{{\rm d} \phantom{y}}{{\rm d} y} \ln \left(P_{n,m}(y)\right)\,=\,\frac{\gamma}{y}\,\frac{y + \tilde{H}_{n,m}(y)}{y-\tilde{H}_{n,m}(y)}\,, \ee
and $\tilde{H}_{n,m}(y)$ satisfies the continuous Painlev{\'e} VI equation
\be  {\cal{P}}_{{\rm VI}}\left(y,\tilde{H}_{n,m}(y);\frac{n^2}{2},-\,\frac{m^2}{2}, \frac{1}{2} (n + m + c)^2 , \frac{1- 4 \gamma^2}{2} \right)\,,\quad c\,\in\,{\mathds{R}}.\ee
\end{subequations}

\begin{rem}
The continuously invariant solution (\ref{invsolQ1P6}) also satisfies the differential-difference equations
\begin{subeqnarray}\label{iscQ1d0}
\alpha\,\frac{\partial v_{(0,0)}}{\partial \alpha} = n\,\frac{(v_{(1,0)}-v_{(0,0)}) (v_{(0,0)}-v_{(-1,0)})}{v_{(1,0)}-v_{(-1,0)}}\,, \label{iscQ1d01} \\ 
\beta\,\frac{\partial v_{(0,0)}}{\partial \beta} = m\,\frac{(v_{(0,1)}-v_{(0,0)}) (v_{(0,0)}-v_{(0,-1)})}{v_{(0,1)}-v_{(0,-1)}}\,. \label{iscQ1d02}
\end{subeqnarray}
The latter are but the invariant surface conditions (\ref{parconf1}), (\ref{parconf2}). On the other hand, the function $v_{(0,0)}$ also satisfies the differential equation
\be
\alpha\,\partial_\alpha v_{(0,0)}\,+\,\beta\,\partial_\beta v_{(0,0)}\,=\,2\,\gamma \,v_{(0,0)}\,, \label{iscQ1d0a}
\ee
since it is invariant under the symmetry generator ${\bf z}_1 + 2 \gamma {\bf z}_3$. Using equations (\ref{iscQ1d0}) to replace the derivatives of $v_{(0,0)}$ appearing in the last equation, we conclude that every continuously invariant solution must satisfy the following constraint
$$n\,\frac{(v_{(1,0)}-v_{(0,0)}) (v_{(0,0)}-v_{(-1,0)})}{v_{(1,0)}-v_{(-1,0)}}\,+\,m\,\frac{(v_{(0,1)}-v_{(0,0)}) (v_{(0,0)}-v_{(0,-1)})}{v_{(0,1)}-v_{(0,-1)}}\,=\,2\,\gamma\,v_{(0,0)}.$$
Reductions of the Schwarzian KdV equation constructed on the basis of this constraint were presented in \cite{NijP6}. \hfill $\Box$
\end{rem}

\section{Continuous invariant solutions and generating equations} \label{SecSud}

The notion of {\it{generating partial differential equations}} was introduced by Nijhoff, Joshi and Hone in \cite{NHJ}, where their archetypical example, the RPDE, was also presented. In the present section, we show that system $\Sigma[u]$ corresponding to several members of the ABS class is intimately related to the above kind of equations. Our method of deriving this relation enables us to produce the results of \cite{NHJ} in a more systematic way, as well as to extend these results to other integrable lattice equations. In particular, we show that equations {\it H1}--{\it H3} and {\it Q1} are related to RPDE.

\begin{rem}
In order to simplify the resulting expressions, in the present section, we adopt the following notation for the corresponding $u_{(i,j)}$
$$u\,=\,u_{(0,0)}\,, \quad u_1\,=\,u_{(1,0)}\,,\quad u_2\,=\,u_{(0,1)}\,.$$
\hfill $\Box$
\end{rem}

\subsection{The system $\Sigma[u]$ of {\it H1}, {\it H2} and {\it Q1}}

Let $S(U,u_1,u_2;\delta)$ denote the following system of partial differential equations:
\begin{subequations} \label{Q1sys}
\begin{eqnarray}
\frac{\partial u_1}{\partial \beta} &= & \frac{u_1-u_2}{\alpha-\beta} \, \left(m\,-\,(u_1-u_2)\,\frac{\partial U}{\partial \beta}\right) \,+\,(\alpha-\beta)\,\delta^2\,\frac{\partial U}{\partial \beta}\,, \label{Q1sys1}\\
& & \nonumber\\
\frac{\partial u_2}{\partial \alpha} &= & \frac{u_1-u_2}{\alpha-\beta} \, \left( n \,+\,(u_1-u_2)\,\frac{\partial U}{\partial \alpha} \right) \,-\,(\alpha-\beta)\,\delta^2\,\frac{\partial U}{\partial \alpha}\,, \label{Q1sys2} \\
& & \nonumber\\
\frac{\partial^2 U}{\partial \alpha \partial \beta} &=& \frac{1}{\alpha-\beta} \left( 2\,(u_1-u_2) \frac{\partial U}{\partial \alpha} \frac{\partial U}{\partial \beta}\,+\,n \frac{\partial U}{\partial \beta}\,-\,m \frac{\partial U}{\partial \alpha}\right)\,. \label{Q1sys3}
\end{eqnarray}
\end{subequations}
One arrives at the above system starting from $\Sigma[u]$ corresponding to equations {\it H1}, {\it H2} and {\it Q1} through the following contact transformation
\begin{equation}\label{Q1miura}
{\cal{M}}(u,U)\,:=\,\left\{ 
\begin{array}{c}
u_{,\alpha} \,=\, \frac{2 \, f(u,u_1,\alpha)\, U_{,\alpha}\, +\,  n\, f_{,u_1}(u,u_1,\alpha)}{2\,r(\alpha)}\\ 
\\ 
u_{,\beta} \,=\, \frac{2 \, f(u,u_2,\beta)\, U_{,\beta}\, +\,  m\, f_{,u_2}(u,u_2,\beta)}{2\,r(\beta)}
\end{array}
\right. \,.
\end{equation}
In particular, ${\cal{M}}(u,U)$ maps
\begin{enumerate}[1.]
\item $\Sigma[u]$ of {\it H1} to $S[u;0]$,
\item $\Sigma[u]$ of {\it H2} to $S[U;\delta]$ with $\delta^2 = 1$, and 
\item $\Sigma[u]$ of {\it Q1} to $S[U;\delta]$.
\end{enumerate}

In view of these observations, system $S[U;\delta]$ introduced above incorporates the continuously invariant solutions of the three integrable lattice equations {\it H1}, {\it H2} and {\it Q1}. What is remarkable is the fact that $S[U;\delta]$ is also related to RPDE, which has been shown to be a generating equation for the KdV hierarchy \cite{NHJ}.

Indeed, $S[U;\delta]$ can be decoupled leading to a fourth order partial differential equation for each of the functions involved. To see this, we first solve equation (\ref{Q1sys3}) for the difference $u_1-u_2$ to find
\be u_1\,-\,u_2\,=\,\frac{1}{2}\,\left((\alpha-\beta) \frac{U_{,\alpha \beta}}{U_{,\alpha} U_{,\beta}}\,+\,\frac{m}{U_{,\beta}}\,-\,\frac{n}{U_{,\alpha}} \right)\,. \label{intrpde1} \ee
Substituting the above expression into (\ref{Q1sys1}), (\ref{Q1sys2}), we obtain $\partial_\beta u_1$ and $\partial_\alpha u_2$ in terms of $U$ and its derivatives. Then, we differentiate equation (\ref{intrpde1}) with respect to $\alpha$ and use (\ref{Q1sys2}) and (\ref{intrpde1}) to eliminate $\partial_\alpha u_2$ and $u_1-u_2$, respectively. This gives $\partial_\alpha u_1$ in terms of the derivatives of $U$. The compatibility between the resulting expression and the first equation of $S[U;\delta]$ leads to the following fourth order partial differential equation 
\be {\cal{R}}(\alpha,\beta,U;n,m)\,+\,2\, \delta^2\, \frac{\partial U}{\partial \alpha} \,\frac{\partial U}{\partial \beta} \,\left(2 \frac{\partial^2 U}{\partial \alpha \partial \beta} - \frac{1}{\alpha-\beta} \left(\frac{\partial U}{\partial \alpha}-\frac{\partial U}{\partial \beta} \right) \right)\,=\,0\,,\label{dRPDE} \ee
where
\begin{eqnarray*}
 {\cal{R}}(\alpha,\beta,U;n,m) &:=& -U_{,\alpha \alpha \beta \beta} +  U_{,\alpha \alpha \beta} \left(\frac{1}{\alpha-\beta} + \frac{U_{,\beta \beta}}{U_{,\beta}} + \frac{U_{,\alpha \beta}}{U_{,\alpha}} \right) + U_{,\alpha \beta \beta} \left(\frac{1}{\beta-\alpha} + \frac{U_{,\alpha \alpha}}{U_{,\alpha}}+ \frac{U_{,\alpha \beta}}{U_{,\beta}} \right)\\
 && - U_{,\alpha \alpha} U_{,\beta \beta} \frac{U_{,\alpha \beta}}{U_{,\alpha} U_{,\beta}} 
 + \, U_{,\alpha \alpha} \left(\frac{n^2}{(\alpha-\beta)^2} \frac{U_{,\beta}^2}{U_{,\alpha}^2} - \frac{1}{\alpha-\beta} \frac{U_{,\alpha \beta}}{U_{,\alpha}} - \frac{U_{,\alpha \beta}^2}{U_{,\alpha}^2} \right) \\
 && + U_{,\beta \beta} \left(\frac{m^2}{(\alpha-\beta)^2} \frac{U_{,\alpha}^2}{U_{,\beta}^2} + \frac{1}{\alpha-\beta} \frac{U_{,\alpha \beta}}{U_{,\beta}} - \frac{U_{,\alpha \beta}^2}{U_{,\beta}^2} \right) \\
&& + \, \frac{n^2}{2 (\alpha-\beta)^3} \frac{U_{,\beta}}{U_{,\alpha}} \left( U_{,\alpha} + U_{,\beta} + 2 (\beta-\alpha) U_{,\alpha \beta}\right)  \nonumber \\
 && -\frac{m^2}{2 (\alpha-\beta)^3} \frac{U_{,\alpha}}{U_{,\beta}} \left( U_{,\alpha} + U_{,\beta} + 2 (\alpha-\beta) U_{,\alpha \beta}\right) + \frac{1}{2 (\alpha-\beta)} U_{,\alpha \beta}^2 \left(\frac{1}{U_{,\alpha}} - \frac{1}{U_{,\beta}}\right)\,. \label{Rpde}
\end{eqnarray*}

When $\delta=0$, the last equation reduces to ${\cal{R}}(\alpha,\beta,U;n,m) = 0$, and this is exactly the equation named RPDE \cite{NHJ,TTX1}. In fact, even when $\delta \ne 0$, equation (\ref{dRPDE}) is essentially the same to RPDE. Specifically, one only needs to set
$$   \tilde{U} \,=\, \frac{1}{2\,\delta}\,\exp\left( 2 \delta U \right)\,,$$
in order to transform equation (\ref{dRPDE}) to ${\cal{R}}(\alpha,\beta,\tilde{U};n,m) = 0$. For this reason, equation (\ref{dRPDE}) will be referred to as RPDE for all values of parameter $\delta$.

The function $U$ may also be considered as a potential for $u_1$ and $u_2$. To see this, we just have to solve $S[U;\delta]$ for the derivatives of $U$. The compatibility of the resulting equations leads to the following system for $u_1$, $u_2$:
\begin{subequations} \label{Q1u1u2sys}
\begin{eqnarray}
 {u_1}_{,\alpha \beta} &=& \frac{2 (u_1-u_2)}{(u_1-u_2)^2-\delta^2 (\alpha-\beta)^2}\,\left({u_1}_{,\alpha}\,{u_1}_{,\beta}\,+\,2\,\delta^2\,(n+1)\,m \right) \nonumber \\
 \nonumber \\
& &- \,\left(\frac{2 \delta^2 (\alpha-\beta)}{(u_1-u_2)^2-\delta^2 (\alpha-\beta)^2}\,+\,\frac{1}{\alpha-\beta} \right)\,\left(m\,{u_1}_{,\alpha}\,+\,(n+1)\,{u_1}_{,\beta}\right)\,, \label{Q1u1u2sys1} \\
 \nonumber \\
 {u_2}_{,\alpha \beta} &=& \frac{2 (u_2-u_1)}{(u_1-u_2)^2-\delta^2 (\alpha-\beta)^2}\,\left({u_2}_{,\alpha}\,{u_2}_{,\beta}\,+\,2\,\delta^2\,n\,(m+1) \right)\nonumber \\
 \nonumber \\
&&+ \, \left(\frac{2 \delta^2 (\alpha-\beta)}{(u_1-u_2)^2-\delta^2 (\alpha-\beta)^2}\,+\,\frac{1}{\alpha-\beta} \right)\,\left((m+1)\,{u_2}_{,\alpha}\,+\,n\,{u_2}_{,\beta}\right) \,. \label{Q1u1u2sys2} 
\end{eqnarray}
\end{subequations}

When $\delta=0$, the last equations decouple easily yielding the RPDE pair
$${\cal{R}}(\alpha,\beta,u_1;n+1,m) = 0\,,\quad {\cal{R}}(\alpha,\beta,u_2;n,m+1) = 0\,.$$
In case $\delta \ne 0$, system (\ref{Q1u1u2sys}) may also be decoupled but the resulting equations are much more complicated. Specifically, one may solve equation (\ref{Q1u1u2sys1}) for $u_2$ and substitute the result into equation (\ref{Q1u1u2sys2}). This leads to a fourth order, second degree partial differential equation for $u_1$, which is omitted here because of its length. Analogous considerations hold for the function $u_2$.

\begin{rem}
System $S[U;0]$ first appeared in \cite{NHJ}. A generalization of $S[U;0]$ was derived in \cite{TTX1,TTX2} in the context of a symmetry reduction of the anti self dual Yang Mills equations. The relation of the latter to the Ernst-Weyl equation and the Painlev{\'e} transcendents were also presented in \cite{TTX2}.  \hfill $\Box$
\end{rem}

\begin{rem}
As already noted, $S[U;\delta]$ is integrable. A Lax pair for this system is given by
\begin{subequations}
\begin{eqnarray}
 \Psi_{,\alpha} = \frac{1}{\alpha-\lambda}\,\left( \begin{array}{c c} n + u_1 U_{,\alpha} & -\,U_{,\alpha} \\  u_1\,(n + u_1 U_{,\alpha}) & -u_1 \,U_{,\alpha}  \end{array} \right)\,\Psi\,-\,\left(\begin{array}{cc} 0 & 0 \\ \delta^2 (\alpha-\lambda)\,U_{,\alpha} & 0 \end{array} \right)\,\Psi\,,\\
\nonumber\\
 \Psi_{,\beta} = \frac{1}{\beta-\lambda}\,\left( \begin{array}{c c} m + u_2 U_{,\beta} & -U_{,\beta} \\ u_2\,(m + u_2 U_{,\beta})& -u_2\,U_{,\beta}   \end{array} \right)\,\Psi\,-\,\left(\begin{array}{cc} 0 & 0 \\ \delta^2 (\beta-\lambda)\,U_{,\beta} & 0 \end{array} \right)\,\Psi\,.
\end{eqnarray}
\end{subequations}
It can be obtained from the Lax pair (\ref{laxsigma}) using the transformation ${\cal{M}}$ and performing the gauge transformation
$$\Phi \,=\,(\alpha-\lambda)^{-n/2}\,(\beta-\lambda)^{-m/2} \,\Psi\,.$$
\hfill $\Box$

\end{rem}

\begin{rem}
It is worth mentioning that, equation (\ref{dRPDE}) is the Euler - Lagrange equations
$$\frac{\partial^2 {\phantom{\partial \alpha}}}{\partial \alpha \,\partial \beta}\,\left( \frac{\partial {\cal{L}}}{\partial U_{,\alpha \beta}}\right)\,-\,\frac{\partial {\phantom{\alpha}}}{\partial \alpha}\,\left( \frac{\partial {\cal{L}}}{\partial U_{,\alpha}}\right)\,-\,\frac{\partial {\phantom{\beta}}}{\partial \beta}\,\left( \frac{\partial {\cal{L}}}{\partial U_{,\beta}}\right)\,=\,0 $$
corresponding to the Lagrangian
$${\cal{L}}\,=\,\frac{\alpha-\beta}{2}\,\frac{U_{,\alpha \beta}^2}{U_{,\alpha}\,U_{,\beta}}\,+\,\frac{1}{2\,(\alpha-\beta)}\,\left(m^2\,\frac{U_{,\alpha}}{U_{,\beta}}\,+\,n^2\,\frac{U_{,\beta}}{U_{,\alpha}} \right)\,+\,2\,\delta^2\,(\alpha-\beta)\,U_{,\alpha}\,U_{,\beta}\,.$$
For $\delta = 0$ this reduces to the Lagrangian for the RPDE given in \cite{NHJ}. \hfill $\Box$
\end{rem}

\subsection{The system $\Sigma[u]$ of {\it H3}}

The continuously invariant solutions of {\it H3} are also related to solutions of RPDE. We establish this connection for the cases $\delta=0$ and $\delta \ne 0$ separately. In each case, we introduce a potential function through a system of equations and use the latter to simplify the corresponding system $\Sigma[u]$. The resulting system can be decoupled leading to the RPDE for the potential function.

\subsubsection{Case I: $\delta = 0$}

Let us first introduce a potential $\psi$ for system $\Sigma[u]$ corresponding to ${\it H3}_{\delta=0}$. This is determined by the relations
\begin{equation} \label{potH3d0}
\psi_{,\alpha}\,=\,\frac{{\mathrm{e}}^U\,\left(n\,-\,\alpha\,U_{,\alpha}\right)}{2\,u_1}\,,\qquad \psi_{,\beta}\,=\,\frac{{\mathrm{e}}^U\,\left(m\,-\,\beta\,U_{,\beta}\right)}{2\,u_2}\,,
\end{equation}
where
$$\exp \left(-\, U(\alpha,\beta) \right) \,=\, u(\alpha,\beta) \,.$$

We solve the above equations for $u_1$, $u_2$ and substitute the resulting expressions into the two first equations of $\Sigma[u]$. Then, we perform the change of the independent variables
$$(\alpha,\,\beta)\,\longrightarrow\,\left(\alpha^2,\,\beta^2\right)$$
and arrive at the following system for $U$ and $\psi$:
\begin{subeqnarray} \label{sysH3fUd0}
 U_{,\alpha \beta} &=& \frac{1}{4\,(\alpha\,-\,\beta)}\,\left(\frac{4\,\alpha^2\, U_{,\alpha}^2\,-\,n^2}{\alpha}\, \frac{\psi_{,\beta}}{\psi_{,\alpha}}\, -\, \frac{4\,\beta^2\, U_{,\beta}^2\,-\,m^2}{\beta}\, \frac{\psi_{,\alpha}}{\psi_{,\beta}} \right), \label{sysH3fUd01} \\
 \nonumber \\
 \psi_{,\alpha \beta} &=& \frac{2}{\alpha\,-\,\beta}\, \left( \alpha\,U_{,\alpha}\, \psi_{,\beta}\, -\, \beta\,  U_{,\beta}\, \psi_{,\alpha} \right)\,. \label{sysH3fUd02}
\end{subeqnarray}
On the other hand, using the above substitutions for $u$, $u_1$ and $u_2$ and the change of the independent variables, the third equation of $\Sigma[u]$ is identically satisfied by taking into account system (\ref{sysH3fUd0}).

The pair of equations (\ref{sysH3fUd0}) can be decoupled, and this leads to the following equation for the potential $\psi$:
$${\cal{R}}(\alpha,\beta,\psi;n,m)\,=\,0\,. $$
The decoupling can be achieved by solving equation (\ref{sysH3fUd02}) for one of the first order derivatives of $U$, e.g. $U_{,\alpha}$, and taking the compatibility condition between the resulting equation and (\ref{sysH3fUd01}). The result is a relation for $U_{,\beta \beta}$. Finally, the compatibility condition between the latter and (\ref{sysH3fUd01}) implies that $\psi$ satisfies RPDE.

On the other hand, system (\ref{sysH3fUd0}) may be decoupled leading to a fourth order, second degree partial differential equation for function $U$. First, we solve (\ref{sysH3fUd01}) for $\psi_{,\alpha}$ to get
$$\psi_{,\alpha}\,=\,\psi_{,\beta}\,A\,,\qquad \psi_{,\alpha \beta}\,=\,\psi_{,\beta}\,B\,,$$
where
$$A\,=\,\frac{2\,\alpha\,\beta\,(\alpha\,-\,\beta)\, U_{,\alpha \beta}\,+\,X}{\alpha\,(m^2\,-\,\beta^2\,U_{,\beta}^2)}\,,\qquad B\,=\,\frac{2\,(\alpha\, U_{,\alpha}\,-\,\beta \,A\, U_{,\beta})}{\alpha\,-\,\beta}\,, $$
and
$$X\,=\,\sqrt{\alpha\,\beta\,\left(4\,\alpha\,\beta\,(\alpha \,-\,\beta )^2\, U_{,\alpha \beta}^2\,+\,(m^2\,-\,\beta^2\,U_{,\beta}^2)\, (n^2\,-\,\alpha^2\,U_{,\alpha}^2) \right)}\,. $$

The compatibility condition $\partial_{\beta} \psi_{,\alpha} = \psi_{,\alpha \beta}$ implies
$$ \psi_{,\beta \beta}\,=\,\psi_{,\beta}\,\left( \frac{B \,-\,{\mathrm{D}}_\beta\,A}{A} \right)\,. $$
Finally, the compatibility condition $\partial_\beta\,\psi_{,\alpha \beta} = \partial_\alpha\,\psi_{,\beta \beta}$ leads to
$${\mathrm{D}}_\alpha\,{\mathrm{D}}_\beta\,\ln \,A\,=\,{\mathrm{D}}_\alpha\,\left( \frac{B}{A}\right)\,-\,{\mathrm{D}}_\beta\,B\,. $$

If we write out the last equation explicitly, solve it for $X$ and square the result, then we end up with a fourth order, second degree (in the highest derivative $U_{,\alpha \alpha \beta \beta}$) partial differential equation. It is the {\it{modified partial differential equation}} (MPDE) presented in \cite{NHJ}.

\subsubsection{Case II: $\delta \ne 0$}

In this case, we introduce the potential $\phi$ by the relations
$$\phi_{,\alpha}\,=\,\frac{u_1 \,(\, n \,u\, +\, \alpha \,u_{,\alpha}\,)}{\alpha\, (\,u \,u_1\, +\, \delta\, \alpha\,)}\,,\quad \phi_{,\beta}\,=\,\frac{u_2 \,(\, m\, u\, +\, \beta\, u_{,\beta}\,)}{\beta\, (\,u\, u_2\, +\, \delta\, \beta\,)}\,,$$
in view of which, $u_1$, $u_2$ may be expressed in terms of $u$, $\phi$ and their derivatives:
\begin{equation} \label{potH3dne0}
 u_1\,=\,\frac{\delta\,\alpha^2\,\phi_{,\alpha}}{\alpha\,u_{,\alpha}\,+\,u\,(n\,-\,\alpha\,\phi_{,\alpha})}\,, \qquad u_2\,=\,\frac{\delta\,\beta^2\,\phi_{,\beta}}{\beta\,u_{,\beta}\,+\,u\,(m\,-\,\beta\,\phi_{,\beta})}\,.
\end{equation}

We substitute the above relations into the two first equations of $\Sigma[u]$ and set
\be 
\begin{array}{l}
u(\alpha,\beta)\,:=\,\exp\left(-\,U(\alpha,\beta)\,+\,\delta\,\psi(\alpha,\beta)\right)\,,\\
\\
\phi(\alpha,\beta) \,:=\,-\,U(\alpha,\beta)\,-\,\delta\,\psi(\alpha,\beta)\, +\,n\,\ln \alpha \,+\,m\,\ln \beta \,. \end{array}  \label{potH3dne0a} \ee
Finally, the change of the independent variables
\be (\alpha,\,\beta)\,\longrightarrow \, \left(\alpha^2,\,\beta^2\right) \label{potH3dne0b}\ee
leads to the following system:
\begin{subeqnarray} \label{sysH3ff}
U_{,\alpha \beta} &=& \frac{1}{4 (\alpha-\beta)} \left(\frac{4 \alpha^2 U_{,\alpha}^2-n^2}{\alpha}\, \frac{\psi_{,\beta}}{\psi_{,\alpha}}- \frac{4 \beta^2 U_{,\beta}^2-m^2}{\beta} \frac{\psi_{,\alpha}}{\psi_{,\beta}}\right)  +\delta^2\psi_{,\alpha} \psi_{,\beta} , \label{sysH3ff1} \\
  \nonumber \\
\psi_{,\alpha \beta} &=& \frac{2}{\alpha\,-\,\beta}\, \left( \alpha\,U_{,\alpha}\, \psi_{,\beta}\, -\, \beta\,  U_{,\beta}\, \psi_{,\alpha} \right) \,.\label{sysH3ff2}
\end{subeqnarray}
Moreover, the third equation of $\Sigma[u]$ is satisfied identically, in view of transformations (\ref{potH3dne0})--(\ref{potH3dne0b}) and by taking into account system (\ref{sysH3ff}).

We may decouple system (\ref{sysH3ff}) following the procedure described in the previous subsection. In this fashion, we conclude that, $\psi$ satisfies RPDE, as well.

\begin{rem}
A Lax pair for system (\ref{sysH3ff}) is given by
\begin{eqnarray}
 \Psi_{,\alpha} \,=\, \frac{1}{\alpha\,-\,\lambda^2}\,\left(\begin{array}{c c}
\alpha\,U_{,\alpha}\,+\,\lambda^2\,\delta\,\psi_{,\alpha} & 2 \,\lambda\,\psi_{,\alpha}\\
\frac{\lambda}{8\,\alpha\,\psi_{,\alpha}}\,\left(n^2\,-\,4\,\alpha^2\,\left(U_{,\alpha}\,+\,\delta\,\psi_{,\alpha}\right)^2 \right) & -\,\alpha\,U_{,\alpha}\,-\,\lambda^2\,\delta\,\psi_{,\alpha}
\end{array} \right)\,\Psi\,, \nonumber\\
 \label{H3laxRM}\\
 \Psi_{,\beta} \,=\, \frac{1}{\beta\,-\,\lambda^2}\,\left(\begin{array}{c c}
\beta\,U_{,\beta}\,+\,\lambda^2\,\delta\,\psi_{,\beta} & 2 \,\lambda\,\psi_{,\beta}\\
\frac{\lambda}{8\,\beta\,\psi_{,\beta}}\,\left(m^2\,-\,4\,\beta^2\,\left(U_{,\beta}\,+\,\delta\,\psi_{,\beta}\right)^2 \right) & -\,\beta\,U_{,\beta}\,-\,\lambda^2\,\delta\,\psi_{,\beta}
\end{array} \right)\,\Psi\,. \nonumber
\end{eqnarray}
The above equations follow from (\ref{laxsigma}) by making the transformations (\ref{potH3dne0})--(\ref{potH3dne0b}) and, subsequently, performing the gauge transformation
$$\Phi \,=\, \left( \begin{array}{c c}
\exp\left(\frac{\delta\,\psi\,-\,U}{2}\right) & 0 \\
0 & \exp\left(\frac{U\,-\,\delta\,\psi}{2}\right)
\end{array} \right)\,\Psi \,.$$
\hfill $\Box$
\end{rem}

\subsection{Connection with previous results} \label{sec-conne}

The preceding analysis shows that the integrable lattice equations {\it H1}--{\it H3} and {\it Q1} are closely related, i.e. their continuously invariant solutions may be expressed in terms of solutions of RPDE. 

The relation between {\it H1}, {\it H3}${}_{\delta=0}$ and {\it Q1}${}_{\delta=0}$ and RPDE was presented by Nijhoff, Hone and Joshi in \cite{NHJ}, starting from a different point of view. Specifically, the authors presented systems of differential--difference equations compatible with the above lattice equations, which, from our point of view, are the invariant surface conditions (\ref{parconf1},\ref{parconf2}). Using the differential--difference equations, they constructed compatible systems of partial differential equations, which in turn lead to RPDE and MPDE. Actually, the systems appearing in the analysis of Nijhoff, Hone and Joshi do not differ from what we called $\Sigma[u]$.

To clarify this correspondence further, let us point out that,
\begin{enumerate}[i)]
\item The system presented in \cite{NHJ} in relation with {\it H1} is actually $S(u,-u_1,-u_2;0)$.

\item In relation with {\it Q1}${}_{\delta=0}$, the authors of \cite{NHJ} presented a system of differential equations $\Delta$, which can be decoupled leading to a fourth order partial differential equation, called Schwarzian partial differential equations (SPDE). Here, we have presented the corresponding system $\Sigma[u]$, i.e. system (\ref{Q1d0sysf}), which may be decoupled for each involved function leading to RPDE.

However, system $\Delta$ and the resulting SPDE are related to (\ref{Q1d0sysf}) and RPDE, respectively. Indeed, starting from system (\ref{Q1d0sysf}), we make the change of the dependent variables
$$u_1\,=\,u\,+\,\frac{2\,\alpha\,u_{,\alpha}}{n\,(1-{\tilde{u}}_1)}\,,\qquad u_2\,=\,u\,+\,\frac{2\,\beta\,u_{,\beta}}{m\,(1-{\tilde{u}}_2)}\,, $$
and, consequently, the change of the independent variables
$$(\alpha,\,\beta)\,\longrightarrow\,\left( \frac{1}{\alpha}\,,\,\,\frac{1}{\beta}\right)\,.$$
This procedure leads to system $\Delta$. Moreover, RPDE is mapped to SPDE using the above transformation of the independent variables $\alpha$, $\beta$. 

\item Finally, the authors of \cite{NHJ} also presented the MPDE in relation with {\it H3}${}_{\delta=0}$, and a Miura transformation relating MPDE to RPDE. From our point of view, this Miura transformation is system (\ref{sysH3fUd0}), which is equivalent to $\Sigma[u]$ corresponding to {\it H3}${}_{\delta=0}$.
\end{enumerate}

\section{Conclusions and perspectives}

We have presented symmetry reductions of the Adler, Bobenko, Suris equations using both of the extended three point generalized symmetries admitted by them. Such reductions lead to special similarity solutions, which we named continuously invariant solutions. It was proven that these are determined by a system of partial differential equations, $\Sigma[u]$, which is integrable in the sense that, it admits an auto-B{\"a}cklund transformation and a Lax pair.

The symmetry analysis and the corresponding reductions of system $\Sigma[u]$ associated with the discrete potential and Schwarzian KdV equations led to new interesting results. In particular, it was shown that, the continuously invariant solutions of {\it H1} are determined by solutions of the continuous Painlev{\'e} V and VI equations. Similar results and considerations were also presented with regard to equation ${\it Q1}_{\delta=0}$.

We were also able to reveal the connection of $\Sigma[u]$ to generating equations. In particular, we derived the generating equations to the {\it H1}--{\it H3} and {\it Q1} members of the ABS family. In addition, we showed that, the continuously invariant solutions of  {\it H1}, ${\it H3}_{\delta=0}$ and ${\it Q1}_{\delta=0}$ are related to RPDE, in accordance with the results of Nijhoff, Hone and Joshi in \cite{NHJ}.

The construction of continuously invariant solutions of the other members of the ABS family and, especially, of the master equation {\it Q4} is one of the interesting directions in which the present work can be extended. In addition, more general lattice systems possessing the consistency property can also be analyzed in the framework of continuously invariant solutions and generating equations. The discrete Boussinesq equation \cite{TN1} and the discrete modified Boussinesq equation \cite{TN2} are among the better known systems which can be brought into the above framework.

\section*{Acknowledgments}
P. Xenitidis thanks V. Papageorgiou for useful discussions and for bringing to his attention reference \cite{NijP6}.

\appendix
\section{The characteristic polynomials of the ABS equations} \label{fGk}

\begin{flushleft}
\begin{tabular}{llll}
{\it{H1}} : & $f(u,x,\alpha) = 1$ & $k(\alpha,\beta) = \beta -\alpha$& $G(x,y) = (x-y)^2$  \\
{\it{H2}} : & $f(u,x,\alpha) = 2(u + x + \alpha)$  & $k(\alpha,\beta) = \beta -\alpha$& $G(x,y) = (x-y)^2 - (\alpha-\beta)^2$  \\
{\it{H3}} : & $f(u,x,\alpha) = u x + \alpha \delta$  & $k(\alpha,\beta) = \alpha^2-\beta^2$ & $G(x,y) = (y \alpha - x \beta) (y \beta - x \alpha)$\\
& & & 
\end{tabular}

\begin{tabular}{cl}
{\it{Q1}} : &  $ f(u,x,\alpha) \,= \,((u-x)^2 - \alpha^2 \delta^2)/\alpha \,,\quad k(\alpha,\beta) \,=\, -\alpha \beta (\alpha-\beta)$ \\
& \\
& $G(x,y) = \alpha \beta \left((x-y)^2 - (\alpha -\beta)^2 \delta^2\right)$ \\
& \\
\end{tabular}

\begin{tabular}{cl}
{\it{Q2}} : & $ f(u,x,\alpha) = ((u-x)^2 - 2 \alpha^2 (u+x) + \alpha^4)/\alpha\,,\quad k(\alpha,\beta) = -\alpha \beta (\alpha-\beta)$ \\
            & \\
            & $ G(x,y) = \alpha \beta \left((x-y)^2 - 2 (\alpha -\beta)^2 (x+y) + (\alpha-\beta)^4\right)$\\
            & \\
\end{tabular}

\begin{tabular}{cl}
{\it{Q3}} : &  $f(u,x,\alpha) =\frac{1}{4\alpha(\alpha^2-1)} (4 \alpha (\alpha u-x) (\alpha x - u) - (\alpha^2-1)^2 \delta^2)$ \\
            & \\
            & $k(\alpha,\beta) = (\alpha^2 -  \beta^2) (\alpha^2-1) (\beta^2 - 1)$ \\
            & \\
            & $G(x,y) = \frac{(\alpha^2-1) (\beta^2-1)}{4 \alpha \beta} \left(4 \alpha \beta (\alpha y-\beta x) (\beta y-\alpha x)+(\alpha^2-\beta^2) \delta^2\right)$\\
            & \\
\end{tabular}

\begin{tabular}{cl}
{\it{Q4}} : &  $f(u,x,\alpha) = \left((u x+ \alpha (u+x) + g_2/4)^2 - (u+x+\alpha)(4 \alpha u x-g_3) \right)/a $\\
                   & \\
                   & $ k(\alpha,\beta) = \,\frac{a b\left(a^2 b + a b^2 +  \left[12 \alpha \beta^2 - g_2 (\alpha+2\beta) - 3 g_3\right] a + \left[12 \beta \alpha^2 - g_2 (\beta + 2 \alpha) - 3 g_3\right] b \right)}{4 (\alpha-\beta)}$ \\
                   & \\
                   & $G(x,y) = (a_0 x y + a_1 (x+y) +a_2) (a_2 x y + a_3 (x+y)+a_4)$ \\
                   & $\qquad \qquad \quad - (a_1 x y + \bar{a}_2 y + \tilde{a}_2 x + a_3) (a_1 x y + \bar{a}_2 x + \tilde{a}_2 y + a_3)$
\end{tabular}
\end{flushleft}

\section{Proof of Proposition \ref{propBcB}} \label{proofBcB}

Let $u$ be a continuously invariant solution of the integrable lattice equation
\be Q(u_{(0,0)},u_{(1,0)},u_{(0,1)},u_{(1,1)};\alpha,\beta)\,=\,0 \label{autoBacdiseq2} \ee
and $\tilde{u}$ be constructed in terms of $u$ via the auto-B{\"a}cklund transformation ${\mathds{B}}_d(u,\tilde{u},\lambda)$. The function $\tilde{u}$ is another continuously invariant solution of equation (\ref{autoBacdiseq2}), provided that it satisfies the system
\begin{subeqnarray} \label{parconfut}
 Q(\tilde{u}_{(0,0)},\tilde{u}_{(1,0)},\tilde{u}_{(0,1)},\tilde{u}_{(1,1)};\alpha,\beta)&=&0 \,,\label{parconfequt}\\
r(\alpha)\,\frac{\partial\,\tilde{u}_{(0,0)}}{\partial\,\alpha}\,+\,n\,R(\tilde{u}_{(0,0)},\tilde{u}_{(1,0)},\tilde{u}_{(-1,0)},\alpha) &=& 0\,, \label{parconf1ut} \\
r(\beta)\,\frac{\partial\,\tilde{u}_{(0,0)}}{\partial\,\beta} \,+\,m \,R(\tilde{u}_{(0,0)},\tilde{u}_{(0,1)},\tilde{u}_{(0,-1)},\beta) &=&0\,. \label{parconf2ut}
\end{subeqnarray}
Obviously, the first equation holds, since $\tilde{u}$ is constructed using ${\mathds{B}}_d(u,\tilde{u},\lambda)$. It remains to show that equations (\ref{parconf1ut}), (\ref{parconf2ut}) also hold.

To prove that (\ref{parconf1ut}) holds, we differentiate the first equation of ${\mathds{B}}_d(u,\tilde{u},\lambda)$ with respect to $\alpha$. Then, we use equation (\ref{parconf1}) and its shift in the $n$ direction to substitute $\partial_\alpha u_{(0,0)}$ and $\partial_\alpha u_{(1,0)}$, respectively. Moreover, we use the determining equation for the generator $\tilde{\mathbf{v}}_1$ to substitute the derivative of $Q(u_{(0,0)},u_{(1,0)},\tilde{u}_{(0,0)},\tilde{u}_{(1,0)};\alpha,\lambda)$ with respect to $\alpha$. In terms of the above substitutions, we come up with the following equation
$$\left(Q_{,{\tilde{u}}_{(0,0)}}\,+\,Q_{,{\tilde{u}}_{(1,0)}}\,{\cal{S}}_n \right)\,\left(r(\alpha)\frac{\partial\,{\tilde{u}}_{(0,0)}}{\partial\,\alpha}\,+\,n\,R\left(\tilde{u}_{(0,0)},\tilde{u}_{(1,0)},\tilde{u}_{(-1,0)},\alpha \right) \right)\,=\,0\,, $$
where we have omitted the arguments of the function $Q(u_{(0,0)},u_{(1,0)},\tilde{u}_{(0,0)},\tilde{u}_{(1,0)};\alpha,\lambda)$.

Finally, we eliminate the value $u_{(1,0)}$ from the above equation and the latter becomes
$$  \left( G\left(u_{(0,0)},\tilde{u}_{(1,0)}\right) \,+\,h\left(u_{(0,0)},\tilde{u}_{(0,0)}\right)\,{\cal{S}}_n \right)
\left(r(\alpha)\frac{\partial\,{\tilde{u}}_{(0,0)}}{\partial\,\alpha}\,+\,n\,R\left(\tilde{u}_{(0,0)},\tilde{u}_{(1,0)},\tilde{u}_{(-1,0)},\alpha \right) \right)\,=\,0\,.$$

This equation involves the values of the function $\tilde{u}$ and the value $u_{(0,0)}$ through the polynomials $h$, $G$. Thus, the corresponding coefficients of the various powers of $u_{(0,0)}$ must be identically zero. The matrix of the resulting algebraic system has rank two \cite{TTX}, and the system admits only the zero solution, i.e.
\begin{subequations}
\be r(\alpha)\frac{\partial\,{\tilde{u}}_{(0,0)}}{\partial\,\alpha}\,+\,n\, R\left(\tilde{u}_{(0,0)},\tilde{u}_{(1,0)},\tilde{u}_{(-1,0)},\alpha \right)\,=\,0\,. \label{conutilde1}\ee

In the same fashion, we differentiate the second equation of the auto-B{\"{a}}cklund transformation with respect to $\beta$ and use equation (\ref{parconf2}) and the determining equation for the symmetry generator $\tilde{\mathbf{v}}_2$ to get that $\tilde{u}$ also satisfies
\be r(\beta)\,\frac{\partial\,{\tilde{u}}_{(0,0)}}{\partial\,\beta} \,+\,m \,R({\tilde{u}}_{(0,0)},{\tilde{u}}_{(0,1)},{\tilde{u}}_{(0,-1)},\beta)\,=\,0\,.\label{conutilde2}\ee
\end{subequations}
Thus, function $\tilde{u}$ satisfies system (\ref{parconfut}), i.e. it is another continuously invariant solution. \hfill $\Box$

\section{Proof of Proposition \ref{propBTSigma}} \label{proofBTSigma}

Proposition \ref{propBcB} implies that, if $u$ is a continuously invariant solution, then function $\tilde{u}$, determined by ${\mathds{B}}_d(u,\tilde{u},\lambda)$, will be another solution of the same kind, and conversely. In other words, the functions $u$, $\tilde{u}$ satisfy systems (\ref{parconf}) and (\ref{parconfut}), respectively. Thus, we can express the derivatives of $\tilde{u}_{(0,0)}$ in terms of the values and the corresponding derivatives of the function $u$ and conversely.

To achieve this, we solve
\be Q(u_{(-1,0)},u_{(0,0)},\tilde{u}_{(-1,0)},\tilde{u}_{(0,0)};\alpha,\lambda)\,=\,0 \label{Qundertilde} \ee
for $u_{(-1,0)}$ and (\ref{parconf1ut}) for $\tilde{u}_{(-1,0)}$. In terms of the above substitutions, the fraction $1/(u_{(1,0)}-u_{(-1,0)})$ becomes
$$\left(\frac{r(\alpha)}{n}\,\partial_\alpha \tilde{u}_{(0,0)}\,-\,\frac{f_{,\tilde{u}_{(1,0)}}(\tilde{u}_{(0,0)},\tilde{u}_{(1,0)},\alpha)}{2} \right)\,\frac{Q_{,u_{(1,0)}}}{Q_{,\tilde{u}_{(1,0)}} f(\tilde{u}_{(0,0)},\tilde{u}_{(1,0)},\alpha)}+ \frac{Q_{,u_{(1,0)} \tilde{u}_{(1,0)}}}{Q_{,\tilde{u}_{(1,0)}}}\,,$$
where we have omitted the arguments of $Q(u_{(0,0)},u_{(1,0)},\tilde{u}_{(0,0)},\tilde{u}_{(1,0)};\alpha,\lambda)$.

The derivatives of $Q$ involved in the above expression are determined by the relations
\begin{eqnarray*}
&& \frac{Q_{,u_{(1,0)}}}{Q_{,\tilde{u}_{(1,0)}}} = \frac{k(\alpha,\lambda) f(\tilde{u}_{(0,0)},\tilde{u}_{(1,0)},\alpha)}{G(u_{(1,0)},\tilde{u}_{(0,0)},\alpha,\lambda)}\,,\\
 \\
&& \frac{Q_{,u_{(1,0)} \tilde{u}_{(1,0)}}}{Q_{,\tilde{u}_{(1,0)}}} = \frac{1}{2}\,\left( \frac{k(\alpha,\lambda) f_{,\tilde{u}_{(1,0)}}(\tilde{u}_{(0,0)},\tilde{u}_{(1,0)},\alpha)}{G(u_{(1,0)},\tilde{u}_{(0,0)},\alpha,\lambda)}\,+\,\frac{G_{,u_{(1,0)}}(u_{(1,0)},\tilde{u}_{(0,0)},\alpha,\lambda)}{G(u_{(1,0)},\tilde{u}_{(0,0)},\alpha,\lambda)} \right)\,,
\end{eqnarray*}
which hold in view of the equation $Q(u_{(0,0)},u_{(1,0)},\tilde{u}_{(0,0)},\tilde{u}_{(1,0)};\alpha,\lambda)=0$. 

We arrive at equation (\ref{conBac1}) by substituting the above expressions into (\ref{parconf1}) and solving the resulting equation for $\partial_\alpha \tilde{u}_{(0,0)}$. 

Conversely, we solve (\ref{Qundertilde}) for $\tilde{u}_{(-1,0)}$ and (\ref{parconf1}) for $u_{(-1,0)}$. Then, we substitute the resulting expressions into (\ref{parconf1ut}) and solve this equation for $\partial_\alpha u_{(0,0)}$. The final result is identical to the equation obtained by interchanging $u$ and $\tilde{u}$ in (\ref{conBac1}).

Equation (\ref{conBac2}), i.e. the fourth equation of ${\mathds{B}}_c(u,\tilde{u},\lambda)$, can be derived in a similar manner using equations 
$Q(u_{(0,-1)},u_{(0,0)},\tilde{u}_{(0,-1)},\tilde{u}_{(0,0)};\beta,\lambda)=0$ and (\ref{parconf2ut}).

Since the class of continuously invariant solutions is closed under ${\mathds{B}}_d$ and the initial solution $u$ satisfies $\Sigma[u]$, the same holds for the function $\tilde{u}$, i.e. the latter satisfies $\Sigma[\tilde{u}]$. Thus, ${\mathds{B}}_c(u,\tilde{u},\lambda)$ defines an auto-B\"{a}cklund transformation of system $\Sigma[u]$. \hfill$\Box$

\end{document}